\documentclass[prb,superscriptaddress,twocolumn,floatfix]{revtex4-1}

\usepackage[FIGTOPCAP]{subfigure}
\usepackage[usenames]{color}
\usepackage[pdftex]{hyperref,graphicx}
\usepackage{amsmath,amssymb,amsthm,amscd,latexsym,epsfig,bm,dsfont}
\hypersetup{colorlinks = true, urlcolor = blue, linkcolor = blue, citecolor = blue}

\usepackage{ulem} 
\usepackage[dvipsnames]{xcolor}
\usepackage{dsfont}

\newcommand{\zz}{\mathbb{Z}_2}
\newcommand{\z}{\mathbb{Z}}

\def\cC{\mathcal{C}}

\def\f2{{\mathbb F}_2}

\begin{document}
\title{Faithful derivation of symmetry indicators: A case study for topological superconductors with time-reversal and inversion symmetries}
\author{Sheng-Jie Huang}
\affiliation{Condensed Matter Theory Center and Joint Quantum Institute, University of Maryland, College Park, MD 20742, USA}
\author{Yi-Ting~Hsu}
\email{yhsu2@nd.edu}
\affiliation{Condensed Matter Theory Center and Joint Quantum Institute, University of Maryland, College Park, MD 20742, USA}
\affiliation{Department of Physics, University of Notre Dame, Notre Dame, IN 46556, USA}
\date{\today}

\begin{abstract}
Topological crystalline superconductors have attracted rapidly rising attention due to the possibility of higher-order phases, which support Majorana modes on boundaries in $d-2$ or lower dimensions. 
However, although the classification and bulk topological invariants in such systems have been well studied, it is generally difficult to faithfully predict the boundary Majoranas from the band-structure information due to the lack of well-established bulk-boundary correspondence. 
Here we propose a protocol for deriving symmetry indicators that depend on a minimal set of necessary symmetry data of the bulk bands and can diagnose boundary features.  
Specifically, to obtain indicators manifesting clear bulk-boundary correspondence, we combine the topological crystal classification scheme in the real space and a twisted equivariant K group analysis in the momentum space. The key step is to disentangle the generally mixed strong and weak indicators through a systematic basis-matching procedure between our real-space and momentum-space approaches. We demonstrate our protocol using an example of two-dimensional time-reversal odd-parity superconductors, where the inversion symmetry is known to protect a higher-order phase with corner Majoranas. Symmetry indicators derived from our protocol can be readily applied to ab initio database and could fuel material predictions for strong and weak topological crystalline superconductors with various boundary features.  
\end{abstract}
\maketitle

\section{Introduction}
Topological crystalline superconductors (TCsc)\cite{Shiozaki2014,Ando2015,Chiu2016,Trifunovic2017,Geier2018,Khalaf2018,Wang2018,WTe2HOTsc,Zhu2019,Yan2019,Ahn2020,Zhang2020dirac,ModelRXZ,Dinhduy2020} have attracted rapidly rising attention since certain crystalline symmetries can enlarge the classifications and protect new types of topological superconductors. 
In particular, there exists TCsc that belong to the higher-order superconductors, which are $d$-dimensional superconductors that do not support Majorana boundary modes in $d-1$ dimension, but support Majoranas on $(d-2)$ or lower-dimensional boundaries\cite{Trifunovic2017,Geier2018,Khalaf2018,Wang2018,WTe2HOTsc,Zhu2019,Yan2019,Ahn2020,ModelRXZ,Dinhduy2020}. 
One interesting example is that the inversion symmetry can protect a two-dimensional (2D) higher-order superconductor, which hosts Majorana zero modes localized on opposite corners\cite{Khalaf2018,Geier2018,WTe2HOTsc}.

Nonetheless, material realizations of topological superconductors in general, including the higher-order and other types of TCsc with non-trivial Majorana boundaries, are rare to date. 
To narrow down the candidates in vast material databases, a useful tool would be the symmetry indicators\cite{Ono2019,Fischer2020,Geier2020,Ono2020refined} that take superconducting-state or normal-state band structures as input and can predict the type of boundary Majoranas. 
Deriving such a boundary diagnostic requires the knowledge of the necessary set of band structure properties, normally the symmetry eigenvalues of occupied bands on certain parts of the Brillouin zone, and a clear mapping from the bulk topology in the momentum space to the boundary consequence in the real space.

Although there exist plenty prior works on symmetry indicators for topological insulators and superconductors\cite{Po2017,Bradlyn2017,Khalaf2018PRX,Watanabe2018,Watanabe2018AZindicators,ShiozakiIndicator,Ono2019,Po2020,Fischer2020,Geier2020,Ono2020refined,One2020z2}, there is one crucial issue that has not been addressed. 
A symmetry indicator group in principle contains both strong phases protected by the internal and point group symmetries and weak phases protected by translational symmetries only.
However, when one tends to write down explicit expressions for the indicators using the band symmetry data at high-symmetry points, strong and weak indicators generally mix with each other. 
The underlying reason is that there is in fact no simple decomposition between the strong and weak phases when spanning the symmetry indicator group using a generic set of bases\cite{Geier2020,Khalaf2018PRX}. 
We point out that such a mixture could cause misinterpretation and confusion when one tries to apply explicit indicator formula to ab initio data or lattice models without knowing the mixing ratio between strong and weak phases for a given indicator. 
More importantly, given that weak and strong phases carry different boundary features, 
symmetry indicators suffering from such mixtures fail to serve as boundary diagnostics and thus have limited application to actual material searches.
Such an issue has been raised in prior work for a certain case\cite{Geier2020}, but a protocol that can be systematically generalized to general space groups is still absent.

In this work, we provide a double-pronged approach to solve this issue systematically: The first is a real-space classification scheme called the topological crystal formalism\cite{Song2017,Huang2018,shiozaki2018generalized,Shiozaki2019,Song2019TC,freed2019invertible,Song2020,Song2020a}, 
and the second is a momentum-space classification scheme, where we perform the equivariant K theory analysis\cite{Freed2013,Shiozaki2017Ktheory,Kruthoff2017,shiozaki2018atiyahhirzebruch,stehouwer2018classification} using a formalism called the Atiyah-Hirzebruch Spectral Sequence (AHSS)\cite{shiozaki2018atiyahhirzebruch,stehouwer2018classification}. 
Through the real-space approach, we can easily identify the boundary type for each topological crystalline phase, whereas through the momentum-space approach, we can obtain the minimal set of band symmetry data that contains the full classification information for each symmetry class. Importantly, the momentum-space information is closely tied to the band topology and is often accessible from the experiments or ab initio calculations. 
By establishing the canonical map between the two complementary approaches, we arrive at symmetry indicators that are capable of diagnosing boundary signatures.

The topological crystal scheme is a method that follows from the recent rapid progress on general real-space-based classification methods for topological crystalline phases\cite{Song2017,Huang2018,shiozaki2018generalized,cheng2018rotation,Shiozaki2019,Song2019TC,freed2019invertible,Song2020,Rasmussen2020,Song2020a,Zhang2020}.
The key idea is that any topological phase with crystalline symmetries is adiabatically connected to a topological crystal state, which is a real-space crystalline pattern of topological states. One can then classify topological crystalline phases in terms of topological crystals, whose classifications can be obtained by taking the well-known Altlan-Zirnbauer (AZ) classes as building blocks and performing a series of systematic assembly processes. 
Since the topological crystal approach is a real-space-based method with a transparent physical picture, in most cases we readily know the boundary signatures. 
Importantly, a nice feature of the topological crystal approach is that the strong and weak phases have a simple decomposition characterized by a set of real-space topological invariants.

As for the momentum-space-based classification method, the main tool for classifying the topological crystalline phases is the equivariant K theory\cite{Freed2013,Shiozaki2017Ktheory,Kruthoff2017,shiozaki2018atiyahhirzebruch,stehouwer2018classification}. 
The calculation of the equivariant K group is in general a difficult task but an important progress has been achieved by using the Atiyah-Hirzebruch Spectral Sequence (AHSS)\cite{shiozaki2018atiyahhirzebruch,stehouwer2018classification}. 
The rough idea is to first decompose the full K group into local contributions from the high-symmetry points, lines, and planes etc., then study how these local contributions assemble together without closing the gap over the Brillouin zone. 
Once the equivariant K group is obtained, we can then derive the corresponding symmetry indicators by quotienting out the subgroup of atomic insulators or superconductors. 
However, there exists an ambiguity in the basis choice when spanning the group of symmetry indicators, which prevents a simple decomposition for the strong and weak phases. 
By bridging the topological crystal approach with this K theory analysis, we provide a systematic protocol to identify the canonical bases for the symmetry indicator group, in which each indicator is purely strong or purely weak and thus corresponds to unambiguous boundary types.

To explicitly demonstrate our protocol, we take two-dimensional class-DIII superconductors with inversion and translational symmetries (in wallpaper group $\bf{p_2}$) as a case study throughout the work. 
We choose to study this class of superconductors because this is arguably the simplest case where the crystalline symmetry protects a higher-order superconducting phase supporting Majorana corner modes. Moreover, it has been shown that such a phase could be realized by introducing odd-parity pairing to a quantum-spin-Hall system\cite{WTe2HOTsc}. This general recipe together with a previously proposed material candidate monolayer WTe2\cite{WTe2HOTsc} that satisfies the recipe provide ample testing models for our protocol and the resulting symmetry indicators.

The rest of the paper is organized as follows. 
In section II, we first review the topological crystal approach and explain how to obtain the real-space classification as well as the boundary modes for each of the topologically distinct phases. We then perform the analysis for our case study, and propose a set of real-space topological invariants based on the topological crystal construction.   
In section III, we first review how to compute the twisted K group for a system with a given symmetry group with crystalline symmetries. We then calculate the K group for our case study and show that the result matches with that from the complimentary real-space approach. 
In section IV, we combine the real-space and k-space results to derive a set of k-space-based symmetry indicators for our case study that matches with our real-space topological invariants. Specifically, this ``basis-matching'' procedure we propose allow us to arrive at symmetry indicators that reflect the building block construction, and are therefore the corresponding boundary type. 
Finally in section V, we test our symmetry indicators for the case study against tight-binding models for various 2D inversion-symmetric superconductors in which the boundary types were readily known from existing numerical or analytical studies. These examples include models with momentum-independent and -dependent inversion operators, as well as both minimal models and ab-initio-based models for monolayer WTe2. 

\section{Real-space perspective: Topological crystal approach}
In this section, we will apply the topological crystal approach on our system: two-dimensional (2D) odd-parity superconductors made out of spinful electrons with time-reversal and wallpaper-group $\bf{p_2}$ symmetries, which consists of an inversion and two translation symmetries $T_x$, $T_y$ in $x$ and $y$ directions. To be precise, the symmetry group $G$ of the Bogoliubov de Genne (BdG) Hamiltonian describing such 2D superconductors is specified by the following relations:
\begin{eqnarray}
\Theta^{2} &=& -1,~\Xi^{2} =1,~\mathcal{I}^{2} = 1 \nonumber
\\
\left[ \Theta,\Xi \right] &=& 0,~\left[ \Theta,\mathcal{I} \right] = 0 ,~\{ \Xi, \mathcal{I} \} = 0,
\label{eqn:g}
\end{eqnarray}
where $\Theta$, $\Xi$, and $\mathcal{I}$ are the time-reversal, particle-hole, and inversion symmetries acting on the BdG Hamiltonian respectively. In particular, the particle-hole and inversion symmetries anticommute because the superconducting order parameter is parity-odd \cite{Ono2020refined,Geier2020}. 

In subsection A, we will demonstrate each step of the topological crystal approach in details for this system and arrive at the real-space classification. In subsection B, we will further define a set of real-space topological invariants based on the building block constructions in the topological crystal analysis. These real-space invariants are boundary indicators that take building-block constructions as inputs.

\subsection{Real-space classification}

Topological crystal method is a unified, real-space approach to classify cSPT phases\cite{Song2017,Huang2018,shiozaki2018generalized,Shiozaki2019,Song2019TC,freed2019invertible,Song2020,Song2020a}. The key idea is that any topological phase with crystalline symmetries is adiabatically connected to a real-space crystalline assembly of topological states. The special states built by such an assembly are dubbed topological crystals. Since the topological crystal approach is a real space based method, it is usually easier to obtain the boundary signature of a cSPT phase.

\begin{figure}
\includegraphics[width=1\columnwidth]{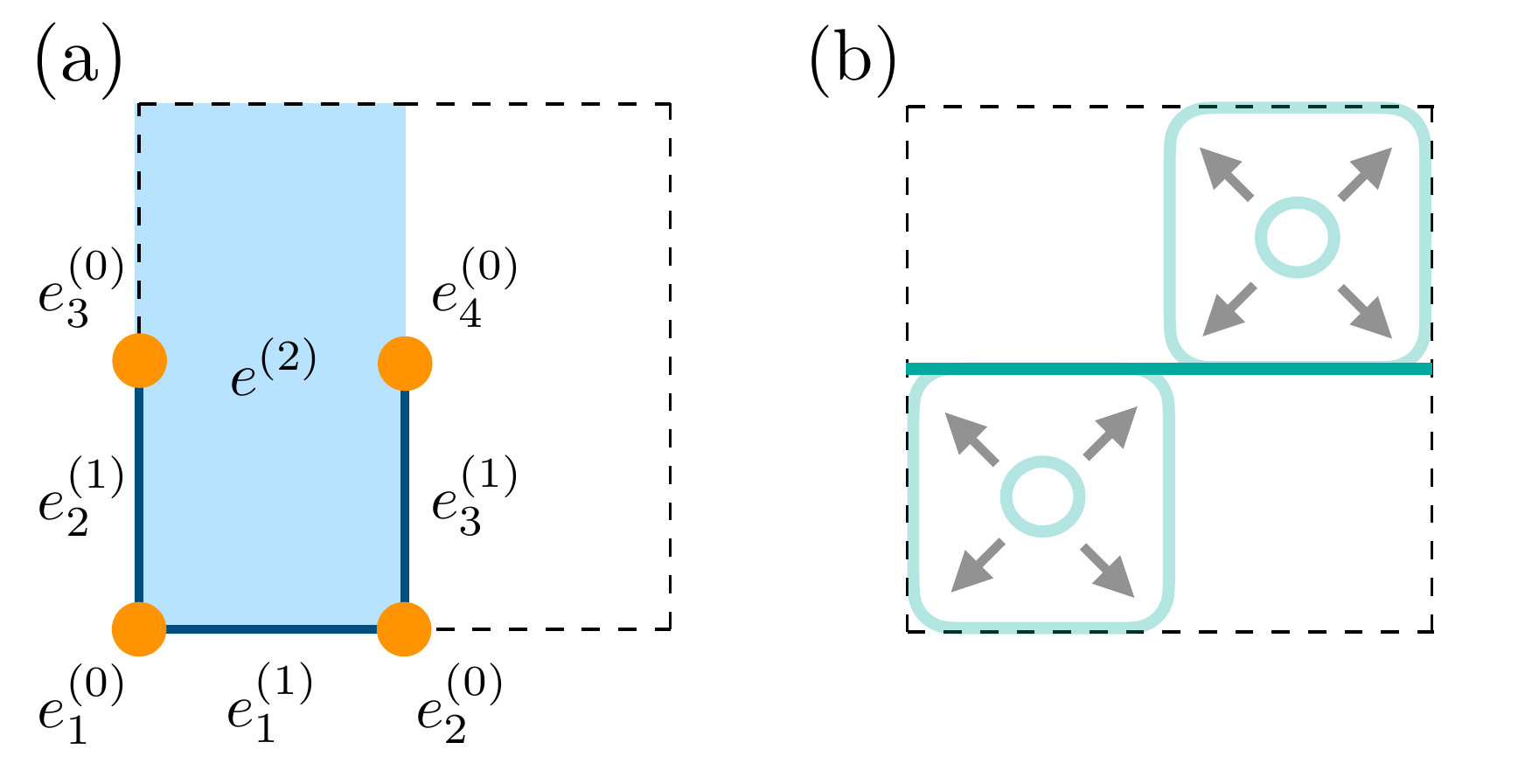}
\caption{(a) The asymmetric unit (AU) and cell decomposition for wallpaper group $\bf{p_2}$. Only one unit cell (bounded by dashed lines) is shown. A $2$-cell is placed in AU, which is represented by the filled blue region. Solid blue lines represents the $1$-cells. Orange dots represents $0$-cells, which are inversion-invariant points. The inversion center is chosen to be at the origin labeled by $e^{(0)}_{1}.$ (b) A weak topological crystal protected by $t_{y}$ translation built by placing a $\hat{x}$-direnctional 1dTSC at $y=1/2$ (darker green line) with real space topological invariants $(\delta_{\{\bar{1}|00\}}$, $\delta_{\{1|10\}}$, $\delta_{\{1|01\}}) = (0,0,1)$ (see main text for the definition). This state is adiabatically connected to a $d_{b}=1$ topological crystal built by placing 1dTSCs on $1$-cells $e^{(1)}_{1}$, $e^{(1)}_{2}$, $e^{(1)}_{3}$ (lighter green lines) through a bubble equivalence relation.
}
\label{fig:real_cells}
\end{figure}

With a 2D superconductor preserving the symmetries in $G$ in mind, we now perform the first step of the approach, which is to define a cell decomposition for the 2D real space. We start by identifying an asymmetric unit (AU), which is the largest interior of a real-space region in which no two points are related by a crystalline symmetry. 
For wallpaper group $\bf{p_2}$, we choose the AU to be the area bounded by $0 < x < a_x/2$ and $0 < y < a_y$ with $a_{x/y}$ being the lattice constants, then copy this AU throughout the 2D space using the translational and inversion symmetries. With such a choice of AU, we now define a cell decomposition for the 2D space as follows: The two-dimensional AUs are dubbed \textit{2-cells}, the one-dimensional edges where two neighboring $2$-cells meet are dubbed \textit{1-cells}, and the zero-dimensional points where four neighboring $1$-cells meet are dubbed \textit{0-cells} [see Fig.~\ref{fig:real_cells}(a)]. While these $0$-cells are located at the inversion-invariant point and their orbit, we require that no two distinct points in the same $1$-cell are related under any action of the crystalline symmetries such that a single $1$-cell exhibits no spatial symmetries\footnote{It is always possible to choose a set of $1$-cells satisfying this property by dividing up the $1$-cells until the condition is satisfied.}.

The second step is to figure out what kind of topological state can be placed on each individual cell. 
Specifically, we place a $d_{b}$-dimensional topological state on each $d_{b}$-cell, where we define the the block dimension $d_{b}$ = 0, 1, 2. These topological states are dubbed the ``$d_{b}$-dimensional building blocks", and the 2D systems constructed by these $d_{b}$-dimensional building blocks are dubbed the $d_{b}$-dimensional topological crystals. Since we define the cell decomposition in a way that none of the cells exhibit crystalline symmetries on its own, the building blocks are topological states protected by internal symmetries and are thus classified by the AZ classification. 

The third step is to ``assemble'' the $2$-cells, $1$-cells, and $0$-cells back together and study how the classifications for 2D, 1D, and 0D building blocks change. 
Specifically, we consider two types of assembly processes, one accounts for the effects on a $d_b$-dimensional topological crystal from lower-dimensional cells with dimension $d_1<d_b$, and the other accounts for the effects from higher-dimensional cells with dimension $d_2>d_b$. 

The former is called the \textit{gluing process}, where we glue two $d_b$-dimensional blocks together and examine whether there exists inevitable gapless modes on the $d_1$-dimensional interface that cannot be gapped out through any symmetry-preserving perturbations.
The latter is to consider the so-called \textit{bubble equivalence relations}. These are equivalence relations between $d_b$-dimensional topological crystal states with and without ``bubbles'' of topological states in neighboring $d$-dimensional cells with $d>d_b$. 
Specifically, suppose we have $d_{b}$-dimensional building blocks that can be glued together at all interfaces with dimension $d_{1} < d_{b}$. We consider a process where bubbles of $q$-dimensional topological states with $q \geq d_{b}$ are locally created in the neighboring $d_2$-dimensional cells. These bubbles can be adiabatically expanded and merge with $d_b$-dimensional building blocks in a symmetry-preserving way. Since this is a symmetry-allowed adiabatic process, 
the new topological crystal with the presence of bubbles is topologically equivalent to the original topological crystal.  
Therefore, to obtain the effective classifications after the assembly processes, we need to consider the modification to the naive classifications of the building blocks due to both the bubble equivalence relations and the glueing processes.


In the following, we will compute the classifications $\mathcal{C}_{d_{b}}(G)$ for $d_{b}$-dimensional topological crystals with symmetry $G$. We start from the $d_{b}=2$ case. Since there is no crystalline symmetry in a $2$-cell, the only possible $d_{b}=2$ building block is the 2D topological superconductor in AZ class DIII (2dTSC), which supports counter-propagating Majorana edge modes and has a $\zz$ classification. We expect that there is no obstruction to glue 2d TSCs together along $1$-cells and $0$-cells since it's known that the 2d TSC is compatible with the inversion symmetry \cite{ShiozakiIndicator,WTe2HOTsc}. The classification thus remains unchanged upon the assembly process. The classification of $d_{b}=2$ topological crystal is therefore $\mathcal{C}_{2}(G) = \zz$. 

Next, we consider the $d_{b}=1$ case. Similarly, since a $1$-cell does not preserve any crystalline symmetry, the only possible $d_{b}=1$ building block that lives on a single $1$-cell is the 1D topological superconductor in AZ class DIII (1dTSC), which supports Majorana Kramers pairs at its two ends and has a $\zz$ classification. 
Given that there are three inequivalent $1$-cells $e^{(1)}_{1}$, $e^{(1)}_{2}$, and $e^{(1)}_{3}$ in each unit cell [see Fig. \ref{fig:real_cells}(a)], the $d_b=1$ topological crystals are classified by $\zz^{3}$ before any assembly process. 
In fact, this classification remains unchanged upon inversion-symmetric gluing and the bubble equivalence operations. This is because we expect that two $d_{b}=1$ blocks can be glued together without producing any gapless mode given that there are known models for 1D class-DIII superconductors that are compatible with the inversion symmetry\cite{ShiozakiIndicator}.
Moreover, the bubble topological states created in the neighboring $2$-cells always come in pairs and trivialize each other. We thus find $\mathcal{C}_{1}(G) = \zz^{3}$. 

Finally, we consider the case of $d_{b}=0$ building blocks, which are simply 0-dimensional BdG Hamiltonians. The resulting $d_{b}=0$ topological crystals and the superconductors that are adiabatically connected to these states can be regarded as ``atomic superconductors'' (ASC), which are superconducting analogue of atomic insulators. We view such ASC as topologically trivial because they do not host topologically protected boundary modes on open geometries. Our definition of ASC is the same as Ref. \onlinecite{ShiozakiIndicator,Geier2020}.


The classification of ASC experiences an AZ-class shift from DIII. This is because the $0$-cells are located at inversion-invariant points [see Fig. \ref{fig:real_cells}(a)] so the $d_{b}=0$ building blocks enjoy an additional on-site $\zz$ symmetry besides the time-reversal and the particle-hole symmetries. 
We find the effective AZ class under this $\zz$ symmetry to be AII due to the following reason: 
In the presence of this additional $\zz$ symmetry, we can always block-diagonalize a $d_{b}=0$ building block, i.e. a 0-dimensional BdG Hamiltonian, into an even-parity block and an odd-parity block. 
However, since the inversion symmetry $\mathcal{I}$ commutes with the time-reversal symmetry $\Theta$ but anticommutes with the particle-hole symmetry $\Xi$, each block is invariant under only $\Theta$, but not $\Xi$.
The effective AZ class of each block is thus AII instead of DIII so that the classification is $\z$ for a 0D building block. 
Given that there are four inversion-invariant points in a unit cell, the classification for ASC is $\z^4$ before considering any assembly processes. 

There are two kinds of bubble equivalence relations that we need to consider for the $d_{b}=0$ building blocks. First, we consider bubbles of zero-dimensional states created in the $1$-cells and $2$-cells. Since the 0-dimensional states in the $1$-cells and $2$-cells are in AZ class DIII, which has a trivial classification, the 0-dimensional bubbles are trivial. 

The second kind is the bubble equivalence relation from the one-dimensional states. This kind is also trivial due to the following reason. Suppose we create a bubble of 1dTSC in the $2$-cell and its orbits, and expand these bubbles adiabatically until they meet on the $1$-cells. Since there are always two bubbles of 1dTSCs meeting on the same $1$-cell, they are expected to trivialize each other on that $1$-cell and leave small bubbles of 1dTSCs enclosing the neighboring $0$-cells. Now the question becomes whether the classification of the 0D BdG Hamiltonians on 0-cells will change when these small bubbles of 1dTSCs shrink to points. 
To answer this, we note that a 1dTSC can be thought of as two independent copies of Kitaev chain related by time-reversal symmetry. 
Moreover, it is known that the ground state of a Kitaev chain with anti-periodic boundary condition is vacuum \cite{Kitaev2001} and that 
the odd-pairing pairing in an inversion-symmetric Kitaev chain implies that each chain satisfies anti-periodic boundary condition.  
We therefore expect that the resulting ground state on $0$-cells must be a vacuum state when the small bubbles of 1dTSCs shrink to a point. Consequently, this kind of bubble equivalence relation does not change the classification of $d_{b}=0$ building blocks either, and we find that the classification of $d_{b}=0$ topological crystals remains $\mathcal{C}_{0}(G) = \z^{4}$ after considering all symmetry-allowed assembly processes.


Equipped with the classifications $\mathcal{C}_{d_b}(G)$ for $d_{b}=2,1,0$ topological crystals, we are now ready to study the full classification $\mathcal{C}(G)$ for 2D time-reversal superconductors with odd-parity pairing. 
Importantly, this full classification $\mathcal{C}(G)$ is \textit{not} a simple product of $\mathcal{C}_{0}(G)$, $\mathcal{C}_{1}(G)$, and $\mathcal{C}_{2}(G)$. The reason is that topological crystals with a fixed block dimension $d_{b}>0$ may not always form a group on their own because stacking more than one copies of such states could result in a topological crystal with a lower block dimension.  
In other words, topological crystals of different block dimensions need not be independent of each other, and it requires a non-trivial group extension among $\mathcal{C}_{d_b}(G)$ to arrive at the final $\mathcal{C}(G)$. The general structure is discussed in Ref.~\onlinecite{Huang2018}, and here we briefly review the two-dimensional case to be self-contained. Define ${\cal D}_{d_b\leq d}(G)$ as the classification of topological crystals with block dimension $d_b$ \emph{less than or equal to} $d=0,1,2$. The states in ${\cal D}_{d_b\leq d}(G)$ do form a group even under the usual stacking operation because stacking two states in ${\cal D}_{d_b\leq d}(G)$ does not produce a topological crystal with block dimension higher than $d$. Due to this fact, we naturally have a sequence of subgroups
\begin{equation}
0 \subset {\cal D}_{d=0}(G) \subset {\cal D}_{d\leq1}(G) \subset {\cal D}_{d\leq2}(G),  
\end{equation}
where by definition ${\cal D}_0(G) = \cC_0(G)$ and ${\cal D}_2(G) = \cC(G)$. 
Moreover, the groups ${\cal D}_{d_b\leq d}(G)$ and the classifications $\cC_{d_b}(G)$ of the topological crystals are related by
\begin{equation}
\cC_{d_b}(G) = \frac{{\cal D}_{d_b}(G)}{{\cal D}_{d_b - 1}(G)}, 
\end{equation}
which could involve non-trivial group extensions and requires careful checks of relations among states with different block dimensions. 

We now turn to our example of 2D class-DIII odd-parity superconductors. 
We start from ${\cal D}_{d_b\leq 1}(G)$, which can be obtained from 
\begin{equation}
\frac{{\cal D}_{d\leq 1}(G)}{{\cal C}_{0}(G)} = \cC_{1}(G).
\end{equation}
Although we know $\cC_{1}(G) = \zz^{3}$ and ${\cal C}_{0}(G)=\z^{4}$, the group ${\cal D}_{d_b\leq 1}(G)$ may not be the simple product of the two and the answer depends on the state resulting from stacking two 1dTSCs. In the absence of time-reversal symmetry, it has been shown that stacking two 1D Kitaev chains in an inversion-symmetric way results in a leftover 0D complex fermion with even parity at the inversion center. Similarly, for our time-reversal symmetric case, given that one 1dTSC can be thought of as two copies of Kitaev chains related by time-reversal symmetry, we expect that stacking two 1dTSCs leads to a 0D Kramers doublet with even parity at the inversion center. Since stacking two $d_b=1$ topological crystals leads to a $d_b=0$ topological crystal, we find 
\begin{equation}
{\cal D}_{d_b\leq 1}(G) \cong \z^{4}. 
\end{equation}

Similarly, to obtain the group ${\cal D}_{d_b\leq2}(G)$, which is given by 
\begin{equation}
\frac{{\cal D}_{d_b\leq2}(G)}{{\cal D}_{d_b\leq1}(G)} = \cC_{2}(G) 
\end{equation}
with $\cC_{2}(G)=\z_2$ and ${\cal D}_{d_b\leq 1}(G)=\z^{4}$, we need to study the resulting state from stacking two 2dTSCs. 
Specifically, we consider two copies of a minimal model for a 2dTSC, each carries a pair of counter-propagating Majorana edge modes. By systematically writing down symmetry-allowed mass terms that couple the two 2dTSCs, we find that there exists no mass term that can fully gap out the Majorana edge modes, and the mass terms with the least level of spatial modulation still leave two leftover Majorana Kramers pairs related by inversion [see Appendix.~\ref{app:doubleTSC}]. Ref.~\onlinecite{cheng2018rotation, Song2020a} achieved the same conclusion for a closely related symmetry class. These two inversion-related Majorana Kramers pairs imply a non-trivial 1dTSC passing through the inversion center. 
Since stacking two $d_b=2$ topological crystals leads to a $d_b=1$ topological crystal, we find 
\begin{equation}
{\cal D}_{d_b\leq 2}(G)={\cal C}(G) \cong \z^{4}. 
\label{eqn:tsc-classification}
\end{equation}
This is the full real-space classification for all 2D time-reversal superconductors with odd-parity pairing, including the atomic superconductors.  

Nonetheless, for our purpose of deriving boundary diagnostics, it is insightful to quotient out the classification ${\cal C}_{0}(G)$ of ASC from the full classification ${\cal C}(G)$ since ASC do not exhibit topologically protected boundary modes on open geometries. 
By doing so, we find that the resulting classification for $d_b=1$ and $d_b=2$ topological crystals is 
\begin{equation}
{\cal C}_{TSC}(G) = \frac{{\cal C}(G)}{{\cal C}_{0}(G)} \cong \z_{4} \times \zz \times \zz.
\label{eqn:real_indicators}
\end{equation}
Note that this group for topological superconductors ${\cal C}_{TSC}(G)$ with symmetry $G$ is not a simple product of the $d_b=1,2$ topological crystals $\mathcal{C}_{2}(G) = \zz$ and $\mathcal{C}_{1}(G) = \zz^3$ due to the non-trivial group extension.  
In particular, as we will show in the nexts subsection, the $\z_4$ and $\z_2$'s in ${\cal C}_{TSC}(G)$ correspond to strong and weak phases with various boundary signatures. 

\subsection{Boundary modes and real-space topological invariants}

In this subsection, we look closer into each of the phases in the group ${\cal C}_{TSC}(G)$ for $d_b=1,2$ topological crystals and define a corresponding set of real-space topological invariants. 
Specifically, we show what the corresponding building block configuration, the resulting boundary modes, and the corresponding topological invariants are for each phase. 
From these information, we find that ${\cal C}_{TSC}(G)$ contains a $\z_4$ that corresponds to strong phases generated by $d_b=2$ building blocks, and two $\z_2$'s that correspond to weak phases generated by $d_b=1$ building blocks. The weak phases support Majorana bands on partial edges, whereas the strong phases support Majorana modes on the edges and/or at the corners. 

\subsubsection{An invariant for $d_b=2$ topological crystals}
First, we define a real-space topological invariant for topological crystals built by different configurations of $d_b=2$ building blocks (2dTSCs).   
It is well-known that the topological crystal states built by placing odd copies of 2dTSCs on $2$-cells are strong phases that support counter-propagating Majorana edge modes protected by time-reversal symmetry while those built by even copies do not support Majoranas on edges. 
In correspondence to such block constructions, we define $\delta_{d_b=2}$ to be a real-space $\zz$ invariant such that $\delta_{d_b=2}=0$ and $1$ characterize the topological crystal states built by even and odd copies of 2dTSCs on 2-cells, respectively.

\subsubsection{Invariants for $d_b=1$ topological crystals}
Next, we move on to defining the real-space invariants for states built by different configurations of $d_b=1$ building blocks (1dTSCs). In the following, we define three $\zz$ invariants $\delta_g=0,1$, one for each symmetry operation $g$ in the wallpaper group $\bf{p_2}$, following the method proposed in Ref. \onlinecite{Song2019TC}. 
We first choose an AU and let $\boldsymbol{r}$ be an arbitrary point in it. Then for a given building block configuration, we set $\delta_g=0$ and $1$ for each operation $g$ if a path connecting point $\boldsymbol{r}$ to point $g\boldsymbol{r}$ crosses through an even and odd number of 1dTSC on 1-cells, respectively. 
As shown in Ref.~\onlinecite{Song2019TC}, these invariants form a homomorphism from $\bf{p_2}$ to $\zz$ and satisfies $\delta_{g_{1}g_{2}} = \delta_{g_{1}} + \delta_{g_{2}}$. 
Moreover, $\delta_g$ defined in this way is independent of the choice of paths. 

Specifically for wallpaper group $\bf{p_2}$, we denote the set of real-space topological invariants to be 
\begin{equation}
\mathfrak{\Delta}_{d_b=1} = (\delta_{\{\bar{1}|00\}}, \delta_{\{1|10\}}, \delta_{\{1|01\}}).
\label{eqn:delta_db1}
\end{equation}
for a given $d_b=1$ block configuration.
Here, each $\zz$ invariant has the form $\delta_g=\delta_{\{\eta_{I}|\eta_{T_x}\eta_{T_y}\}}$, where $\eta_{I}=1$ and $\bar{1}$ indicates the absence and presence of the inversion protection, and $\eta_{T_{x/y}} =0,1$ indicates the absence and presence of the protection by translation in the $x/y$ direction. 
In particular, $\delta_{\{\bar{1}|00\}}$ is a ``strong" topological invariant, which characterizes an inversion-protected phase. 
The two indices $\delta_{\{1|10\}}$ and $\delta_{\{1|01\}}$ are ``weak" topological invariants---these characterize the weak phases protected by translation symmetries $T_{x}$ and $T_{y}$ respectively. 

We now discuss how the set of invariants $\mathfrak{\Delta}_{d_b=1}$ correspond to various block configurations with concrete examples. 
There are three independent $1$-cells $e^{(1)}_{1}$, $e^{(1)}_{2}$, and $e^{(1)}_{3}$ in an unit cell [see Fig. \ref{fig:real_cells}(a)] where we can place $d_b=1$ building blocks to build different block configurations that preserve inversion and translational symmetries. 
We start from placing 1dTSCs on $1$-cell $e^{(1)}_{3}$ in each unit cell. According to our definition for the real space topological invariant $\mathfrak{\Delta}_{d_b=1}$, the resulting phase is  characterized by $\mathfrak{\Delta}_{d_b=1}=(0,1,0)$. Given such a configuration, we expect the Majorana end modes from these 1dTSCs to form Majorana bands along the $x$-directional edges of an open geometry.  
Moreover, such a $\mathfrak{\Delta}_{d_b=1}=(0,1,0)$ topological crystal state is a weak phase protected by the translational symmetry $T_x$ alone. This is because the 1dTSCs can be trivialized in pairs in an inversion-symmetric way when the $T_x$ translation is broken, but any trivialization is prohibited by $T_x$ even when the inversion symmetry $\mathcal{I}$ and $T_y$ translation are broken.  

Similar to the $\mathfrak{\Delta}_{d_b=1}=(0,1,0)$ phases, a $\mathfrak{\Delta}_{d_b=1}=(0,0,1)$ phase is also a weak phase but protected by the $T_{y}$ translation only. The corresponding topological crystal state can be constructed by placing a 1dTSC along the $x$-direction that passes through $(0,1/2)$ in each unit cell, and consequently supports Majorana bands along the $y$-directional edges. 
We point out that this topological crystal is adiabatically connected through an equivalence operation to the topological crystal built by placing three 1dTSCs per unit cell, one on each of the 1-cells $e^{(1)}_{1}$, $e^{(1)}_{2}$, and $e^{(1)}_{3}$ and their orbit. 
The equivalence operation that connects the two states is to locally create a bubble of 1d TSC in the $2$-cells, and adiabatically grow these bubbles until they meet the neighboring $1$-cells while preserving the symmetry [see Fig.~\ref{fig:real_cells}(b)]. Since 1dTSCs trivialize in pairs, we clearly see that these two topological crystal states are connected to each other by this equivalence operation, and are both labeled by $\mathfrak{\Delta}_{d_b=1}=(0,0,1)$. 

Next, we study the configuration that leads to a higher-order topological superconducting state, which supports two inversion-related 0D Majorana Kramers pairs on the boundary. 
Specifically, such a configuration consists of two 1dTSCs per unit cell on the $1$-cells $e^{
(1)}_{2}$ and $e^{(1)}_{3}$, and the resulting state is characterized by $\mathfrak{\Delta}_{d_b=1}=(1,0,0)$\footnote{The state built by placing two 1dTSCs per unit cell, one on the $1$-cell $e^{(1)}_{1}$ and one passing through $(0,1/2)$, is also characterized by $\mathfrak{\Delta}_{d_b=1}=(1,0,0)$.}.
As indicated by the invariant $\mathfrak{\Delta}_{d_b=1}$, this phase is protected by the inversion symmetry alone, which can be understood in the following way. 
When we break the inversion symmetry while preserving the translations, the two 1dTSCs in each of the unit cells can trivialize each other in a translational symmetric way, and no 1dTSC can survive. 
In contrast, when we break the translational symmetries but preserve the inversion, although most of the 1dTSCs can move towards the inversion center and trivialize their inversion partners in an inversion-symmetric way, there is no way to remove the 1dTSCs on the $1$-cells $e^{(1)}_{2}$ that pass through the inversion center.  
Importantly, the surviving 1dTSC supports one Majorana Kramers pair on each end when placed on a geometry that terminates in the $y$ direction. 

When the inversion and translation symmetries are both present, we expect the inversion-related Majorana pairs to be trapped at opposite corners in an open geometry. This can be understood pictorially as follows. 
Suppose we consider a topological crystal built by placing two 1dTSCs per unit cell on the $1$-cells $e^{(1)}_{2}$ and $e^{(1)}_{3}$, and put it on an open square geometry. On each of the top and bottom edges, there are two Majorana Kramers pairs sitting at $x=0$ and $x=1/2$ in every boundary unit cell, and the total number of Majorana Kramers pairs per edge has to be an odd number due to the inversion symmetry. In general, there can be symmetry-allowed boundary perturbations that couple and dimerize these edge Majorana Kramers pairs. Since these boundary perturbations preserve both the inversion and the translation symmetry $T_x$, it is clear that most of the Majorana Kramers pairs would be dimerized except two pairs trapped at inversion-related corners. 
Therefore, we expect the $\mathfrak{\Delta}_{d_b=1}=(1,0,0)$ phase to support Majorana corner modes at two opposite corners, and is thus a higher-order topological superconducting state.

The last example we discuss is a configuration where we place one 1dTSC per unit cell on the $1$-cell $e^{(1)}_{1}$ or $e^{(1)}_{2}$. Let us choose $e^{(1)}_{2}$ without loss of generality. The resulting topological crystal is characterized by $\mathfrak{\Delta}_{d_b=1}=(1,1,0)$. As indicated by the real-space invariants, we expect this state to be a mixture of an inversion $\mathcal{I}$-protected strong phase and a translation $T_{x}$-protected weak phase. Specifically, when we break $\mathcal{I}$ and keep $T_x$, we end up with a $T_{x}$-protected weak phase since there is only one 1d TSC per unit cell. On the other hand, when we break $T_x$ and keep $\mathcal{I}$, we end up with an inversion-protected strong phase since all 1dTSCs can be trivialized in pairs in an inversion-symmetric way except the 1dTSC that crosses the inversion center.
Such a $\mathfrak{\Delta}_{d_b=1}=(1,1,0)$ phase therefore supports two inversion-protected Majorana Kramers pairs at opposite corners as well as translation-protected Majorana bands along the $x$-directional edges.

\subsubsection{The full set of real-space invariants}
After defining the real-space invariants $\delta_{d_b=2}$ and $\mathfrak{\Delta}_{d_b=1}$ that characterize the $d_b=2$ and $d_b=1$ topological crystal states, respectively, we now combine the two and denote  
\begin{equation}
\mathfrak{\Delta} = (\delta_{d_b=2}, \delta_{\{\bar{1}|00\}}, \delta_{\{1|10\}}, \delta_{\{1|01\}})
\label{eq:rspace_inv}
\end{equation}
to be the full set of real-space topological invariants that characterizes the topological superconductor group ${\cal C}_{TSC}(G)$ defined in Eq.~\ref{eqn:real_indicators}. 

We make two important final remarks about this real-space invariant $\mathfrak{\Delta}$ and the topological superconductor group ${\cal C}_{TSC}(G)$. First, the four indices in $\mathfrak{\Delta}$ are in fact \textit{not} linearly independent due to the dependence between the $d_b=2$ and $d_b=1$ topological crystals. Specifically, we define these indices based on the $d_b=2$ and $d_b=1$ block constructions separately, but two copies of $d_b=2$ topological crystals (i.e. two 2dTSCs) in fact lead to a $d_b=1$ topological crystal with two inversion-related Majorana Kramers pairs on the boundary [see discussion near Eq.~\ref{eqn:tsc-classification} and in Appendix.~\ref{app:doubleTSC}. 
Since the former and the latter states are characterized by $\delta_{d_b=2}=2$ and  $\mathfrak{\Delta}_{d_b=1}=(1,0,0)$, respectively, the real-space invariant $\mathfrak{\Delta}$ in Eq. \ref{eq:rspace_inv} satisfies the relation 
\begin{equation}
\mathfrak{\Delta} = (2,0,0,0) \cong (0,1,0,0). 
\label{eq:rspace_constraint}
\end{equation}

The second remark is that the topological superconductor group ${\cal C}_{TSC}(G)$ obtained by quotienting out the atomic superconductors is essentially the quotient group that we will derive for the symmetry indicators by using the momentum-space approach (K theory formalism).
However, it is important to note that this symmetry indicator group and the real-space topological superconductor group ${\cal C}_{TSC}(G)$ are \textit{only isomorphic to each other}. 
In particular, although the group of topological superconductors ${\cal C}_{TSC}(G)$ and the corresponding real-space invariants in $\mathfrak{\Delta}$ have a simple decomposition between the strong and weak phases, there is generally not true for the symmetry indicators due to the freedom in basis choice. 
Specifically, among all the indicator expressions that are related by basis transformations, only the one in the canonical basis does not mix the strong and weak phases. 
As we will show in the next section, the key step in deriving symmetry indicators that are capable of diagnosing boundary features is therefore to identify the canonical mapping that maps from the real-space invariants in $\mathfrak{\Delta}$ to the space of symmetry indicators.

\section{Momentum-space perspective: twisted equivariant K theory}
In this section, we review the momentum-space approach that we will use to classify our system of 2D class-DIII superconductors with odd-parity pairings. Specifically, we explain how to formulate such a classification problem in the presence of crystalline symmetries in terms of a twisted equivariant K group analysis. 
In subsection A, we discuss the cases with internal symmetries only. In subsection B, we turn to the cases with crystalline symmetries.

\subsection{Free fermions with internal symmetries}
It is well known that the non-interacting gapped fermionic phases in the absence or presence of the time-reversal symmetry $\Theta$, particle-hole symmetry $\Xi$, and the chiral symmetry $\mathcal{C}=\Theta \Xi$, can be classified using K theory. Depending on the whether these intrinsic discrete symmetries square to $\pm 1$ or $0$ (absent), there are ten resulting classes dubbed the Altland-Zirnbauer (AZ) classes. The classification of each class in $d$ spatial dimension is given by the corresponding K group, which is the group of equivalence classes of mappings from a $d$-dimensional Brillouin zone ($d$-torus $T^d$) to the classifying space of symmetry-allowed Hamiltonians. 

To see how K theory provides a natural description for the classification of gapped Hamiltonians, we first discuss the rules we expect from physics ground for identifying microscopically different Hamiltonians as topologically equivalent. 
Specifically, there are two equivalence relations that we follow:
First, any two gapped Hamiltonians $H_{0}$ and $H_{1}$ are considered topologically equivalent and belong to the same class if there exists a symmetry-allowed adiabatic path $H(s)$ in the space of local Hamiltonians such that $H(0) = H_{1}$, $H(1) = H_{2}$, and $H(s)$ remains gapped throughout the path $0 \le s \le1$.
Second, since a gapped system should stay in the same class upon adding any additional trivial degrees of freedom, two Hamiltonians $H_{1}$ and $H_{2}$ are considered topologically equivalent if there exists an adiabatic path that connects $H_{1}' = H_{1} + H_{\text{trivial}}$ and $H_{2}' = H_{2} + \tilde{H}_{\text{trivial}}$, where $H_{\text{trivial}}$ and $\tilde{H}_{\text{trivial}}$ are Hamiltonians from the trivial class. Note that this is true even if we cannot find an adiabatic path connecting $H_1$ and $H_2$ directly.  
A valid mathematic description for the classification should account for the above two equivalence relations. 

These equivalence relations are nicely formulated in Karoubi's formulation of K theory, where the K groups consist of adiabatic paths between different gapped Bloch Hamiltonians. Mathematically, an adiabatic path between two gapped Hamiltonians $H_{1}(\bf{k})$ and $H_{2}(\bf{k})$ at momentum $\bf{k}$ is denoted by a triple $((E, G), H_{1}(\bf{k}),$ $H_{2}(\bf{k}))$, 
where $E$ denotes the space of the Bloch states (or more precisely, the vector bundles) on which $H_{1}$ and $H_{2}$ acts, and $G$ denotes the group of symmetries the Hamiltonians and the path obey. 
To simplify the notation, we will suppress the label $G$ unless it's relevant in the context. 
We can now collect all the paths among the Hamiltonians that are identified as topologically equivalent using the first equivalence relation, and label the corresponding equivalence class by any of the paths. In the following, we will slightly abuse the notation and use $(E, H_{1}, H_{2})$ to label either a path between $H_{1}$ and $H_{2}$ or the equivalence class both $H_1$ and $H_2$ belong to depending on the context. 

For these equivalence classes under the first relation that satisfy a given symmetry group $G$, we can define an addition operation between two triples as their direct sum  
\begin{equation}
(E, H_{1}, H_{2}) + (E', H_{1}', H_{2}') = (E \oplus E', H_{1} \oplus H_{1}', H_{2} \oplus H_{2}').  
\label{eqn:gpop}
\end{equation}
We also define the equivalence class $(E, H, H)$ as the trivial class. 
With these properties defined, we can now formulate the second equivalence relation using the notion of triples. 
Specifically, we can expand the definition of equivalence classes by identifying two classes $(E, H_{1}, H_{2})$ and $(E', H_{1}', H_{2}')$ as equivalent if there exist trivial triples $(E_{a}, H_{a}, H_{a})$ and $(E_{b}, H_{b}, H_{b})$ such that
\begin{align}
(E, H_{1}, H_{2}) + (E_{a}, H_{a}, H_{a}) = (E', H_{1}', H_{2}') + (E_{b}, H_{b}, H_{b}). 
\end{align} 
We use $[E, H_{1}, H_{2}]$ to denote this enlarged equivalence class under the second equivalence relation. 
All the equivalence classes $[E, H_{1}, H_{2}]$ satisfying symmetry group $G$ together form a K group $K_{G}(T^d)$ in the K theory, where the base space is a $d$-dimensional Brillouin zone ($d$-torus $T^d$). 
Such a K group is an Abelian group, where the group addition is defined in Eq. \ref{eqn:gpop} with a change of the bracket type, the identity is given by $[E, H, H]$, and the inverse of a triple $[E, H_{1}, H_{2}]$ is given by $[E, H_{2}, H_{1}]$. Note that in the presence of point group symmetries, there could be more inverse elements other than $[E, H_{2}, H_{1}]$ (due to the module structure in the K group\cite{Shiozaki2017Ktheory}), as we will see later in our case study of inversion symmetry.

Importantly, to associate different equivalence classes of adiabatic paths $[E, H_{1}, H_{2}]$ with topologically distinct classes of gapped phases of matter, we pick a ``trivial'' Hamiltonian as the universal reference Hamiltonian instead of using different reference Hamiltonians $H_{2}$ for different triples. For superconductors, a natural choice for this reference point is a BdG Hamiltonian formed by a vacuum state 
\begin{align}
H_0=\text{diag}(\mathds{1}_{N},-\mathds{1}_{N}),
\label{eqn:vacuum}
\end{align}
where $N$ is the number of normal bands. For the rest of the paper, we will therefore consider triples of the form $[E, H, H_0]$, where $H$ is the BdG Hamiltonian of interest. 
We emphasize that the existence of such a universal reference point is crucial when constructing a symmetry indicator 
that works for both superconductors with momentum-dependent and momentum-independent symmetry operators. 
Finally, the K group formed by the symmetry-allowed equivalence classes $[E, H_{1}, H_{0}]$ gives the classification of gapped phases of matter with given internal symmetries.

\subsection{In the presence of crystalline symmetries}
{
Such a classification scheme, however, needs a generalization when the system also has crystalline symmetry. 
We therefore need a formalism to obtain the classification of topological crystalline phases. 
A general scheme called twisted equivariant K theory\cite{Freed2013,Shiozaki2017Ktheory,Kruthoff2017,shiozaki2018atiyahhirzebruch,stehouwer2018classification} was proposed to classify gapped Bloch Hamiltonians of a given AZ class with additional crystalline symmetries, where the classifications are given by the twisted equivariant K group $^\phi K_G^{\tau,-n}(T^d)$. 
Similar to the original K group, this generalized K group classifies the equivalence classes of mappings from the $d$-dimensional Brillouin zone $T^d$ to the classifying space of Hamiltonians allowed by symmetry group $G$. 
However, this symmetry group $G$ now includes both the internal symmetries from the AZ class and the crystalline symmetries. For real AZ classes, it is convenient to introduce a mod-$8$ integer $n$ to label the AZ classes in the following order: AI, BDI, D, DIII, AII, CII, C, CI. The integer label $n$ is defined by the number of additional effective chiral symmetries (see Ref.~\cite{Shiozaki2017Ktheory,shiozaki2018atiyahhirzebruch,stehouwer2018classification} for a precise definition). For instance, the time-reversal superconductors we focus on in this work belong to class DIII, which is labeled by $n=3$. 
 
We now introduce the twisting data $\phi$ and $\tau$, which are additional parameters that need to be specified for systems with crystalline symmetries\cite{Shiozaki2017Ktheory,shiozaki2018atiyahhirzebruch,stehouwer2018classification}. Specifically, the twisting data are
\begin{enumerate}
\item a homomorphism $\phi: G \rightarrow \zz$, where $\phi(g)=1$ if $g\in G$ is unitary and $\phi(g)=-1$ if $g$ is anti-unitary,   
\item a $U(1)$ phase factor $e^{i \tau_{g_{1},g_{2}}(\boldsymbol{k})}$,  which is given by 
\begin{equation}
U_{g_{1}}(g_{2}\boldsymbol{k}) U(g_{2})(\boldsymbol{k}) = e^{i \tau_{g_{1},g_{2}}(g_{1}g_{2}\boldsymbol{k})} U_{g_{1}g_{2}}(\boldsymbol{k}). 
\end{equation}
Here, $\tau_{g_{1},g_{2}}(\boldsymbol{k})$ is in fact a group $2$ cocycle $\tau \in \mathcal{Z}^{2} (G, C(T^{d},U(1)))$, which takes values in the abelian group $C(T^{d},U(1)_{\phi})$ of $U(1)$-valued functions defined on the Brillouin zone with a group action defined by $e^{i (g \cdot \tau)(\boldsymbol{k})} = e^{i \phi(g) \tau(g^{-1}\boldsymbol{k})}$.
\end{enumerate}
The set of phase factors $\tau$ (usually dubbed a factor system) is non-trivial when $G$ forms a projective representation or when the space group $G$ is non-symmorphic. In our case, we focus on topological superconductors in class DIII with symmorphic wallpaper group $\bf{p_2}$ and odd-parity pairing. 
For the symmetry group we consider, since the time-reversal symmetry has $\Theta^2=-1$ and the inversion symmetry $\mathcal{I}$ anticommutes with the particle-hole symmetry $\Xi$ [see Eq. \ref{eqn:g}],  
the non-trivial phase factors $\tau$ in our case are $\tau(\Theta,\Theta)=-1$ and $e^{i \theta_{\mathcal{I}}} = \tau(\mathcal{I},\Xi) / \tau(\Xi,\mathcal{I}) = -1$. In particular, $e^{i \theta_{\mathcal{I}}}$ is the one-dimensional representation of the inversion symmetry $\mathcal{I}$ carried by the gap function $\Delta(\boldsymbol{k})$. Therefore, our goal is to compute the K group $^\phi K_{G}^{\tau,-3}(T^2)$ with the above-mentioned symmetry group $G$ and twisting data ($\phi$, $\tau$). This twisted equivariant K group will give us the classification of 2D time-reversal odd-parity superconductors with wallpaper group symmetry $\bf{p_2}$. 

\section{Computation of twisted equivariant K group} 
The computation of the twisted equivariant K group $^{\phi}K^{\tau,-n}_{G}(T^d)$ is, however, a difficult task in general. It was nonetheless shown in Ref. [\onlinecite{shiozaki2018atiyahhirzebruch,stehouwer2018classification}] that one can make a successive approximation for $^{\phi}K^{\tau,-n}_{G}(T^d)$ by applying a formalism called Atiyah-Hirzebruch (AH) spectral sequence in the momentum space. 
Here, we sketch the general idea of this method: 
Instead of computing the K group for a $d$-dimensional Brillouin zone directly, we can decompose the Brillouin zone into different dimensional subspaces in a symmetry-preserving way such that the systems living on these subspaces have only local symmetries.  
Since the K groups for these subspaces can either be obtained from the AZ classification or be easily computed, the task boils down to assembling these subspaces properly to obtain the K group for the whole system.
To approximate the full K group, we need to figure out how these subspace K groups change upon 
different assembly processes order by order until the result converges. 
In this section, we will take 2D class-DIII superconductors with $\bf{p_2}$ wallpaper group symmetries and odd-parity pairing as a case study and compute the classification using AH spectral sequence.  

\subsection{$E_1$ page}
\begin{figure}
\includegraphics[width=0.5\columnwidth]{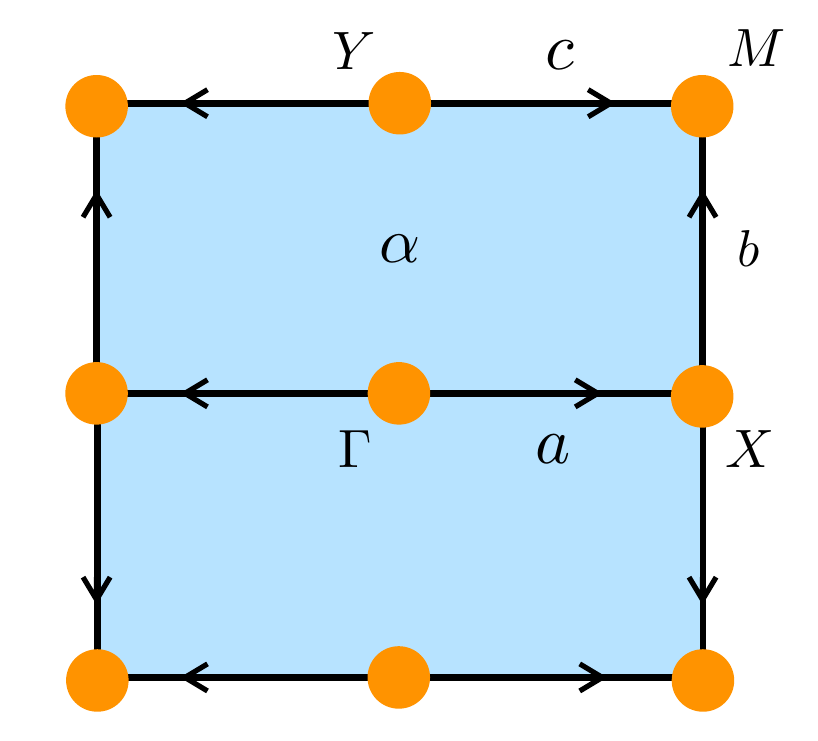}
\caption{The cell decomposition of the Brillouin zone for wallpaper group $\bf{p_2}$. $2$-cells are represented by the blue regions. Block solid lines represents $1$-cells. Orange dots represents $0$-cells  at the high-symmetry points. $\alpha$ labels the independent 2-cell, $a$, $b$, $c$ label the independent 1-cells, and $\Gamma$, $X$, $Y$, $M$ label the independent 0-cells.
}
\label{fig:k_cells}
\end{figure}

We first decompose the 2D Brillouin zone in a symmetry-preserving way into $d_{c}$-dimensional subspaces or ``cells", where $d_{c}=0,1,2$. Similar to the cell-decomposition in real space, we first choose an AU in momentum space. Here we choose AU to be the area bounded by $-\pi/2 < k_{x} < \pi/2$,  $-\pi/2 < k_{y} < \pi/2$. This two-dimensional subspace, or a 2-cell, is labeled by $\alpha$ in Fig.~\ref{fig:k_cells}. 
We then define 1-cells to be the edges where two neighboring 2-cells meet, with the property that no two distinct points in the same 1-cell are related by a spatial symmetry. The 1-cells defined this way are in fact the high-symmetry lines. Similarly, we define 0-cells to be the points where the neighboring 1-cells meet and are located at the inversion-invariant points: $\Gamma$, $X$, $Y$, $M$. These points are also the time-reversal invariant momenta (TRIM). There are three independent 1-cells in the Brillouin [labeled by $a$, $b$, and $c$ in Fig.~\ref{fig:k_cells}], which are also the high-symmetry lines $\Gamma - X$, $X - M$, and $Y - M$, respectively. We further assign orientations for each of the 1- and 2-cells, as shown by the arrows in Fig.~\ref{fig:k_cells}.

Next, we write down the K groups for the sub-systems that live on each cell, which can be done since the sub-systems exhibit only local symmetries. It is convenient to arrange these local K groups, before subjecting to any consistency constraints, into a table dubbed the $E_{1}$-page in the literature\cite{shiozaki2018atiyahhirzebruch,stehouwer2018classification}. 
Specifically, this $E_1$-page table contains local K groups $^{\phi}K^{\tau,-(n-p)}_{G}(X_{p},X_{p-1})$ for the topological phenomena on stand-alone $p$-dimensional sub-systems $X_{p}$ that contains $(p-1)$-dimensional sub-spaces $X_{p-1}$
\begin{equation}
E_{1}^{p,-n} :=^{\phi}K^{\tau,-(n-p)}_{G}(X_{p},X_{p-1}),
\label{eqn:E1df}
\end{equation}
where 
the topological phenomena is in the effective AZ class $(n-p)$ with symmetry group $G$ and symmetry twisting data $(\phi, \tau)$. Note that the remaining topological phenomena associated with the $(p-1)$-dimensional sub-space $X_{p-1}$ is trivial in the relative K group. Moreover, when $p$ is larger than the physical dimension or $n > n_{0}+p$, where $n_{0}$ denotes the physical AZ class, the corresponding entries in the $E_{1}$ page are defined to be trivial. 

For computation convenience, we can further write each entry in the $E_1$ page as a product of the K groups on each of the $p$-cells:
\begin{equation}
E_{1}^{p,-n} = \prod_{j\in\mathbb{D}^p}  {^{\phi_{j}^{p}}K_{G_{j}^{p}}^{\tau_{j}^{p},-(n-p)}}(D_{j}^{p},\partial D_{j}^{p}), 
\label{eqn:K_E1_cells}
\end{equation}
where the cell label $j$ runs over the set of all $p$-cells $\mathbb{D}^p$, and $G_j^{p}$ and $(\phi_{j}^{p},\tau_{j}^{p})$ are the little group and the restricted twisting data for the $j^{th}$ $p$-cell $D_{j}^{p}$.}

The physical meaning of the upper half of the $E_1$-page table was proposed in Ref.~\onlinecite{shiozaki2018atiyahhirzebruch}, and is a direct interpretation of the definition in Eq. \ref{eqn:E1df}. We dub this interpretation the ``topological phenomena interpretation''.
Under such an interpretation, the diagonal entries 
$E_{1}^{p,-p}$ are the classification of gapped Hamiltonians in AZ class 
$p$ on $p$-cells, whereas the first and second upper diagonal entries 
$E_{1}^{p+1,-p}$ and 
$E_{1}^{p+2,-p}$ represent the classifications of the anomalous gapless states on $(p+1)$-cells and singular points on $(p+2)$-cells in the same class, respectively.

\begin{table*}
\begin{center}
\begin{tabular}{ c | c c c c} 
 $n=n_{0}$ & 0D gapped Hamiltonian on 0-cells & Anomalous gapless points on 1-cells & Singular points on 2-cells & {}\ldots{} \\ 
 $n=n_{0}+1$ & ? & 1D gapped Hamiltonian on 1-cells & Anomalous gapless on 2-cells & {}\ldots{} \\ 
 $n=n_{0}+2$ & ? & ? & 2D gapped Hamiltonian on 2-cells & {}\ldots{} \\
  {}\vdots{}    & {}\vdots{}  & {}\vdots{}  & {}\vdots{}  & \\ 
 \hline
 $E_{1}^{p,-n}$ & $p=0$ & $p=1$ & $p=2$ & {}\ldots{} \\
\end{tabular}
\caption{
The $E_{1}$-page of AH spectral sequence. The physical AZ class of the system is denoted by $n_{0}$. The diagonal entries in the $E_{1}$-page are topological invariants of gapped Hamiltonian with AZ class $n_{0}$, which are the fundamental building block of the whole K group. The question marks represent the entries with unknown physical meanings although mathematically they are well-defined.}
\label{table:K-E1_page}
\end{center}
\end{table*}

The physical meaning of the lower half of the $E_1$-page table is, however, not clear under the topological phenomena interpretation [labeled by question marks in Table ~\ref{table:K-E1_page}]. Fortunately, there is an alternative interpretation dubbed the ``representation interpretation'', which can be useful for computation convenience. By invoking an isomorphism, the expression of the $E_1$-page in Eq. \ref{eqn:K_E1_cells} can be written as \cite{shiozaki2018atiyahhirzebruch} 
\begin{eqnarray}
E_{1}^{p,-n} &\cong& \prod_{j \in \mathbb{D}^{p}} {^{\phi_{j}^{p}}K_{G_{j}^{p}}^{\tau_{j}^{p},-n}}(D_{j}^{p})
\\
&\cong& \prod_{j \in \mathbb{D}^{p}} {^{\phi_{j}^{p}}K_{G_{j}^{p}}^{\tau_{j}^{p},-n}}(pt),   
\label{eqn:irrepinterpretation}
\end{eqnarray}
where $pt$ denotes a 0D point in each $p$-cell $D_{j}^{p}$. Under the representation interpretation, which directly follows from Eq. \ref{eqn:irrepinterpretation}, each entry $E_{1}^{p,-n}$ in the table classifies the representations carried by the occupied eigenstates of a class-$n$ 0D gapped Hamiltonian on $p$-cells. 
In the rest of this section, we will adopt these two interpretations at different steps during the calculation depending on which provides more computation convenience. 


Now we compute the $E_1$-page table entry by entry for 2D class-DIII odd-parity superconductors with wall-paper group $\bf{p_2}$ symmetries. Instead of filling out the full table, we will only discuss the entries necessary for computing the full K group, which are $E_{1}^{p,-3}$, $E_{1}^{p,-4}$, and $E_{1}^{p,-5}$ for $p=0,1,2$, except $E_{1}^{2,-3}$, and $E_{1}^{0,-5}$. 

We start with $E_{1}^{0,-3}$, which, according to both the topological phenomena and representation interpretation, is the classification of  representations for a $0$D class-DIII BdG Hamiltonian with odd-parity pairing at the four high-symmetry points $\Gamma$, $X$, $Y$, and $M$. 
Importantly, besides the time-reversal and particle-hole symmetries, there is an extra local $\z_2$ symmetry that comes from the inversion symmetry. Since the high-symmetry points are also inversion-invariant momenta, 
this $\z_2$ symmetry is local and the classification of the 0D BdG Hamiltonian can be calculated in a block-diagonalized form. 
The classification of the 0D BdG Hamiltonian is discussed in detail in Section V A. We find that  the classification is $\z$ for 0D class-DIII Hamiltonians with odd-parity pairing and a $\z_2$ symmetry \footnote{This result was also found in Ref. \onlinecite{Geier2020}}. We therefore find $E_{1}^{0,-3}=\z \times \z \times \z \times \z$ according to Eq. \ref{eqn:K_E1_cells} or Eq.~\ref{eqn:irrepinterpretation} since there are four high-symmetry points.  

We now move to $E_{1}^{1,-3}$ and $E_{1}^{2,-3}$. According to the representation interpretation, these entries are the classifications 
of the representations of a 0D class-DIII inversion-symmetric Hamiltonian at any general points in 1- and 2-cells, respectively. Although the actions of the inversion symmetry $\mathcal{I}$ on 1- and 2-cells are no longer local and brings one momentum point to another, the combined symmetries of $\mathcal{I}$ and the time-reversal $\Theta$ as well as $\mathcal{I}$ and the particle-hole symmetry $\Xi$ are still local in momentum. Since the effective time-reversal and particle-hole symmetries $\tilde{\Theta}=\Theta \mathcal{I}$ and $\tilde{\Xi}=\Xi \mathcal{I}$ have $\tilde{\Theta}^2=\tilde{\Xi}^2=-1$, the effective AZ classes of the Hamiltonians on 1- and 2-cells are both class CII without any extra local symmetry. The classification of such Hamiltonians is trivial\cite{Geier2020}. We therefore have $E_{1}^{1,-3}=E_{1}^{2,-3}=0$. 

The next entry is $E_{1}^{0,-4}$. According to the representation interpretation, this entry classifies the representation of a class-AII inversion-symmetric 0D Hamiltonian placed on 0-cells. Similar to the $E_{1}^{0,-3}$ case, in the presence of the extra local $\z_2$ symmetry, we can find the classification by block-diagonalizing the Hamiltonian in the $\z_2$ space and figure out the effective symmetry class of each block. For each block we find a $\z$ classification\footnote{we write $\z$ instead of $2\z$ for convenience by counting half of the Kramer's doublets.}, which leads to $\z^2$ per high-symmetry point\cite{Geier2020}. We therefore have $E_{1}^{0,-4}=\z^2\times\z^2\times\z^2\times\z^2$ when including all four 0-cells. 

The other two entries for the classifications of 0D class-AII representations are $E_{1}^{1,-4}$ and $E_{1}^{2,-4}$, which are for Hamiltonians on 1- and 2-cells, respectively. By conducting a similar analysis as what we did in the $E_{1}^{1,-3}$ and $E_{1}^{2,-3}$ cases, we find that the effective symmetry classes remain AII for both cases. The classifications are therefore $E_{1}^{1,-4}=\z^2\times\z^2\times\z^2$ and $E_{1}^{2,-4}=\z^2$, given that there are three independent 1-cells and one 2-cell [see Fig.~\ref{fig:k_cells}].  

Finally, for the 0D class-CII representations on 1- and 2-cells, 
we find $E_{1}^{1,-5}=E_{1}^{2,-5}=0$. This is because the effective AZ classes on 1- and 2-cells are both DIII, and the class-DIII classification in 0D is trivial\cite{Geier2020}. This concludes our discussion of the $E_{1}$-page, and the results are summarized in Table \ref{table:E1}. 
\begin{table*}
\begin{center}
\begin{tabular}{ c | c c c c} 
 $n=3$ & $\z \times \z \times \z \times \z$ & $0$ &  \\ 
 $n=4$ & $\z^2\times\z^2\times\z^2\times\z^2$ & $\z\times\z\times\z$ & $\z$  \\ 
 $n=5$ &   & $0$ & $0$ \\ 
 \hline
 $E_{1}^{p,-n}$ & $p=0$ & $p=1$ & $p=2$  \\
\end{tabular}
\caption{Summarized results of the $E_1$-page table. Here we only show the entries that are relevant for obtaining the $E_2$-page.}
\label{table:E1}
\end{center}
\end{table*}

\subsection{Differentials and higher pages}
Now we are ready to ``assemble'' the data at the 0-, 1-, and 2-cells and study the consistency relations that are consequently imposed on the topological phenomena.
In the following, we will discuss the physical meanings of the consistency relations and how such processes lead to the final classification using both the topological phenomena and the representation interpretations. 
Then for computational convenience, we will perform the calculation using the representation interpretation, just as what we did to fill out the $E_1$-page table.

\subsubsection{Physical meanings under the topological phenomena interpretation}
We start with the physical meaning under the topological phenomena interpretation, and take $E_1^{0,-3}$ as an example. According to Table \ref{table:K-E1_page}, $E_1^{0,-3}$ gives the classification of 0D class-DIII gapped Hamiltonians on 0-cells. 
However, this classification on 0-cells should be modified upon the assembly processes where the finite-energy eigenstates on 0-cells classified $E_1^{0,-3}$ 
become part of 1D or 2D bands. 
Such higher-dimensional bands, if not already gapless, could undergo symmetry-allowed band inversions or chemical-potential shifts that inevitably create new gapless points on 1-cells. 
The consistency relation that acts on $E_1^{0,-3}$ is therefore to keep only the 0D gapped Hamiltonians that remain gapped on 1-cells upon extending to higher-dimensional bands.  
Mathematically, we can formulate those symmetry-allowed band inversions or chemical potential shifts by defining the mapping $d_1^{0,-3}: E_1^{0,-3}\rightarrow E_1^{1,-3}$, which is dubbed a first differential. 
Therefore, the consistency relation that amounts to keeping only the gapped 0D Hamiltonians that remain gapped on 1-cells (upon extension to higher-dimensional bands) is imposed by replacing $E_1^{0,-3}$ with the kernel of this mapping $\text{Ker}(d_1^{0,-3})\subseteq E_1^{0,-3}$.

Similar consistency relations arising from assembly processes can also be imposed on 1D and 2D class-DIII gapped Hamiltonians in $E_1^{p,-(3+p)}$, $p=1,2$. 
Generally speaking, for $p$-dimensional gapped Hamiltonians in $E_1^{p,-(3+p)}$, we impose such type of consistency constraints by first defining the first differentials
\begin{align}
d_1^{p,-(3+p)}: E_1^{p,-(3+p)}\rightarrow E_1^{p+1,-(3+p)},
\label{eqn:firstdiff} 
\end{align}
then 
replacing the pre-assembled classification $E_1^{p,-(3+p)}$ by its subset $\text{Ker}(d_1^{p,-(3+p)})$. 
This first type of consistency constraints arises from how 
the classifications of gapped Hamiltonians on $p$-cells (i.e. the three diagonal entries in the $E_1$-page table) are affected by the neighboring higher-dimensional $(p+1)$-cells during the assembly process. 

Besides the first type, we also expect a second type of consistency constraints that results from the effects of lower-dimensional $(p-1)$-cells during the assembly process. 
In principle, such constraints should arise from the processes $d_1^{p-1,-(3+p)}:E_1^{p-1,-(3+p)}\rightarrow E_1^{p,-(3+p)}$, which map the lower off-diagonal entries to the diagonal entries in the same row in the $E_1$-page table. 
In the topological phenomena interpretation, however, the physical meanings of these processes are unclear since the phenomena corresponding to the lower off-diagonal entries are unknown [see the question marks in Table \ref{table:E1}]. 
Fortunately, the representation interpretation, on the other hand, provides clear physical meanings and operational definitions for all the first differentials.
In the next subsection, we will therefore employ the representation interpretation to discuss the meanings of consistency constraints arising from both higher- and lower-dimensional neighboring cells.  

\subsubsection{Physical meanings under the representation interpretation}
In the representation interpretation, where each entry $E_1^{p,-n}$ in the $E_1$-page table represents the classification of class-$n$ representations carried by 0D occupied states located in $p$-cells, the first differential  
$d_1^{p,-(3+p)}$ in Eq. \ref{eqn:firstdiff} represents the process where these representations in $p$-cells split into adjacent $(p+1)$-cells as
\begin{align}
\rho^p_{a,i}=\bigoplus_{b,j}[M_{d_1^{p,-n}}]_{a,i;b,j}\rho^{p+1}_{b,j}, 
\label{eqn:represplit}
\end{align} 
where $\rho^p_{a,i}$ is the representation $a$ on the $i^{\text{th}}$ $p$-cell.
Importantly, since difference dimensional cells have different symmetries, while some linearly independent representations on $p$-cells remain linearly independent on $(p+1)$-cells after such splitting processes, some become redundant. 

We now discuss the two types of consistency constraints arising from the assembly processes for given class-$n$ representations on $p$-cells classified by $E_1^{p,-n}$.  
The first type of constraints states that when assembling $p$- and neighboring $(p+1)$-cells together, we modify the classification $E_1^{p,-n}$ by keeping the representations that are linearly independent only on $p$-cells, but not on $p+1$-cells. These representations are given by $\text{Ker}(d_1^{p,-n})\subseteq E_1^{p,-n}$, which is the same expression as the one we arrive at using the topological phenomena interpretation. 
The second type of constraints, which arises from considering the effects of $(p-1)$-cells on $p$-cells, states that we need to exclude the ``unstable'' representations on $p$-cells that can be trivialized by those split from the adjacent $(p-1)$-cells through the process $d_1^{p-1,-n}:E_1^{p-1,-n}\rightarrow E_1^{p,-n}$. 
Since these unstable $p$-cell representations are described by the image of this mapping $\text{Im}(d_1^{p-1,-n})$, we need to quotient out $\text{Im}(d_1^{p-1,-n})$ from the group classifying the independent representations on $p$-cells.
By taking into account the consistency constraints from both the $(p+1)$- and $(p-1)$-cells, the modified classification is now given by the $E_2$-page  
\begin{align}
E_2^{p,-n}:=\text{Ker}(d_1^{p,-n})/\text{Im}(d_1^{p-1,-n}),  
\label{eqn:E2}
\end{align}
which is the cohomology of the $E_1$-pages.

The $E_2$-page provides the classification for representations on $p$-cells satisfying the consistency relations from the effects of $(p\pm 1)$-cells. However, there are clearly high-order compatibility relations to consider that involve effects from $(p\pm 2)$-cells and beyond during the assembly processes. 
A higher-order process involving $(p\pm r)$-cells with $r>1$, such as a representation splitting process from 0- to 2-cells ($r=2$), 
can be described by an $r^{th}$ differential
\begin{align}
d_r^{p,-n}:E_r^{p,-n}\rightarrow E_r^{p+r,-(n+r-1)}.
\label{eqn:differentials}
\end{align}
With various higher differentials with $r>1$, we can further obtain the higher pages 
\begin{align}
E_{r+1}^{p,-n}:=\text{Ker}(d_r^{p,-n})/\text{Im}(d_r^{p-r+1,-(n-r+2)}),   
\label{eqn:Epages}
\end{align}
which provide the classifications that take into account the higher-order processes involving up to $(p\pm r)$-cells. 

By repeating this procedure until entries in the $E_r$-page table converge, which occurs at $r\leq d$ for $d$-dimensional systems, we obtain the final convergent table dubbed the limiting page $E_{\infty}$. 
The entries in the limiting-page table bear different physical meanings in the two interpretations. 
For instance, in the representation interpretation, 
the limiting pages $E_{\infty}^{p,-(3+p)}$, $p=0,1,2$ are the classification groups for the DIII, AII, and CII representations carried by 0D occupied states on 0-, 1-, and 2-cells respectively, where all the consistency constraints from higher- and lower-dimensional cells are incorporated. 
In contrast, in the topological phenomena interpretation, all three diagonal entries $E_{\infty}^{p,-(3+p)}$, $p=0,1,2$ represent classifications of class-DIII odd-parity superconductors with all consistency constraints considered. Nonetheless, entries with different $p$'s are for superconducting states supported on $p=$0-, 1-, and 2-cells, respectively. 

\subsubsection{The limiting page and the full classification}
The last step is to relate these limiting pages back to the full classification of the 2D topological crystalline phase of our interest. 
Although this relation is not intuitive in the representation interpretation, it is intuitive from the topological phenomena interpretation that we can obtain the classification of the full 2D system by ``stacking'' the sub-system classifications on 0-, 1-, and 2-cells that already incorporated all the consistency constraints. 

In fact, it is mathematically known that the diagonal entries in the limiting-page table $E_{\infty}^{p,-(n+p)}$, $p=0,1,2$ can be used to approximate the relevant twisted equivariant K group on the entire 2D Brillouin zone\cite{shiozaki2018atiyahhirzebruch,stehouwer2018classification} in the following way.   
Let us take the example of 2D class-DIII superconductors with wallpaper group symmetry $\bf{p_2}$ and odd-parity pairings. Here the targeted twisted equivariant K group is $^{\phi}K^{\tau,-3}_{G}(T^2)$, where the symmetry group $G$ and the twisting data ($\phi$, $\tau$) account for all the symmetry properties. To compute the targeted K group through the assembly processes, we can define a series of subgroups $F^{p,-(3+p)}$ for $p=0,1,2$ by  
\begin{equation}
^{\phi}K^{\tau,-3}_{G}(T^{2}) = F^{0,-3}  \supseteq F^{1,-4} \supseteq F^{2,-5}, 
\label{eqn:sequence1}
\end{equation}
such that the limiting pages are given by the quotient groups of these subgroups as
\begin{eqnarray}
E_{\infty}^{p,-(3+p)} &\cong& F^{p,-(3+p)} / F^{p+1,-(3+p+1)}\nonumber
\\
E_{\infty}^{2,-n} &\equiv& F^{2,-n}.
\label{eqn:sequence2}
\end{eqnarray}
Under such a construction, the relations in Eq. \ref{eqn:sequence1} and Eq. \ref{eqn:sequence2} allow us to start from the limiting page $E_{\infty}^{0,-3}$ for the high-symmetry points.  
Then by iteratively incorporating the ones for the high-symmetry lines $E_{\infty}^{1,-4}$ and for general points in Brillouin zone $E_{\infty}^{2,-5}$, we arrive at the full K group for the 2D superconductors with desired symmetries. 
Importantly, the inversion symmetry and the odd-parity pairings are readily implemented in each of the subsystem classifications $E_{\infty}^{p,-(3+p)}$ with appropriate symmetry constraints and consistency relations.\\

\subsubsection{Computation for the case study}
We now apply the above analysis to our case study, and finish calculating the full twisted equivariant K group $^{\phi}K^{\tau,-3}_{G}(T^2)$ for 2D class-DIII odd-parity superconductors with wallpaper group $\bf{p_2}$ symmetries. 
Starting from the $E_1$-page table [see Table \ref{table:E1}] we obtained, we now need to study the consistency constraints from the neighboring higher- or lower-dimensional cells to obtain the higher pages. 
In terms of the actual calculation for the higher pages, it is more convenient to adopt the representation interpretation as we did for the $E_1$-page table 
since the topological phenomena interpretation 
only provides clear meanings for half of the $E_r$-page tables [see Table \ref{table:K-E1_page}]. Nonetheless, for the half of the tables where the topological phenomena interpretation does bear physical meanings, calculations based on the two interpretations should lead to the same results and can serve as a consistency check [see an example in Appendix.~\ref{app:1std}].

We start from the first differentials. In the representation interpretation, as shown in Eq. \ref{eqn:represplit}, a first differential $d_1^{p,-n}$ describe the process of the class-$n$ representations of 0D states on $p$-cells splitting into adjacent $(p+1)$-cells. 
Therefore to calculate the $E_2$-page table defined in Eq. \ref{eqn:E2}, we need to figure out the coefficient matrix $M_{d_1^{p,-n}}$ in Eq. \ref{eqn:represplit}. The coefficients can be positive or negative depending on whether the orientations of the $p$- and $(p+1)$-cells are consistent or inconsistent. 
Since the kernels and images of $d_1^{p,-n}$ represent the representations on $p$-cells that become redundant or remain linearly independent when splitting into ($p+1$)-cells, respectively, practically we can calculate the $E_2$-pages by ``diagonalizing'' the coefficient matrix $M_{d_1^{p,-n}}$ through Smith decomposition.   

Before studying the coefficient matrices, we first list the $E_2$ page entries that do not require actual calculations. These entries can be directly read off from the $E_1$ page due to the adjacent trivial elements in the table [see Table \ref{table:E1}].
Specifically, we have $E_2^{p,-3}=E_1^{p,-3}$ and $E_2^{p,-5}=E_1^{p,-5}$ for $p=0,1,2$. 
We therefore only need to consider the remaining entries $E_2^{0,-4}$, $E_2^{1,-4}$, and $E_2^{2,-4}$ in the following. 

For a class $n=4$ inversion-symmetric system, the 0D occupied states at each of the high-symmetry points $\Gamma$, $X$, $Y$, $M$ can be either parity-even or -odd. Nonetheless, the states on the high-symmetry lines $\Gamma-X$, $Y-M$, and $X-M$ can carry only one single representation since the inversion symmetry acts non-locally. 
The coefficient matrix for the first differential $d_1^{0,-4}:E_1^{0,-4}\rightarrow E_1^{1,-4}$ therefore is a $3\times 8$ matrix 
\begin{align}
M_{d_1^{0,-4}} = \left(\begin{array}{cccccccc}
1 & 1 & -1 & -1 & 0 & 0 & 0 & 0 \\
0 & 0 & 0 & 0 & 1 & 1 & -1 & -1 \\
0 & 0 & 1 & 1 & 0 & 0 & -1 & -1 \\
\end{array}\right)
\end{align}
in the basis of ($\Gamma-X$, $Y-M$, $X-M$) and ($\Gamma_{\text{even}}$, $\Gamma_{\text{odd}}$, $X_{\text{even}}$, $X_{\text{odd}}$, $Y_{\text{even}}$, $Y_{\text{odd}}$, $M_{\text{even}}$, $M_{\text{odd}}$). 
Since the Smith normal form of $M_{d_1^{0,-4}}$ is given by 
\begin{align}
UM_{d_1^{0,-4}}V=
\left(\begin{array}{cccccccc}
1 & 0 & 0 & 0 & 0 & 0 & 0 & 0 \\
0 & 1 & 0 & 0 & 0 & 0 & 0 & 0 \\
0 & 0 & 1 & 0 & 0 & 0 & 0 & 0 \\
\end{array}\right) 
\end{align}
with $U$ and $V$ being the transformation matrices, we find the kernel to be $\text{Ker}(d_1^{0,-4})=\z^5$. 
Physically this means that out of the eight linearly independent 0D representations in $E_1^{0,-4}$, there are five that become redundant when splitting into the neighboring 1-cells due to the lowered local symmetry.  
The image of this mapping can then be obtained through the first isomorphism theorem as $\text{Im}(d_1^{0,-4})=E_1^{0,-4}/\text{Ker}(d_1^{0,-4})=\z^3$. 

The other first differential necessary for computing the $E_2$-pages is $d_1^{1,-4}:E_1^{1,-4}\rightarrow E_1^{2,-4}$. Similar to the above analysis for $d_1^{0,-4}$, we find the coefficient matrix to be a trivial $1\times 3$ matrix 
\begin{align}
M_{d_1^{1,-4}} = (0,0,0)
\end{align}
in the basis of the only 2-cell $\alpha$ and the high-symmetry lines ($\Gamma-X$, $Y-M$, $X-M$). 
The kernel and image of $d_1^{1,-4}$ are therefore given by $\text{Ker}(d_1^{1,-4})=\z^3$ and $\text{Im}(d_1^{1,-4})=0$. 
Following the definition of $E_2$-pages in Eq. \ref{eqn:E2}, 
we find $E_2^{0,-4}=\z^5$, $E_2^{1,-4}=0$, and $E_2^{2,-4}=\z$. 
Equipped with the $E_2$-page table [summarized in Table \ref{table:E2}], we are now ready to proceed to higher differentials and higher pages. 

\begin{table}[]
\centering
\begin{tabular}{ c | c c c c} 
 $n=3$ & $\z \times \z \times \z \times \z$ & $0$ & \\ 
 $n=4$ & $\z^5$ & 0 & $\z$  \\ 
 $n=5$ &  & $0$ & $0$ \\ 
 \hline
 $E_{2}^{p,-n}$ & $p=0$ & $p=1$ & $p=2$  \\
\end{tabular}
\caption{Summarized results for the $E_2$-page table. Here we only show the entries that could be relevant for obtaining the limiting page.}
\label{table:E2} 
\end{table}

We now consider how the $E_2$-page table is further modified by the assembly processes involving representations on $(p\pm 2)$-cells. These processes are described by the second differentials $d_2^{p,-n}$ with $p=0,1,2$ and $n=3,4,5$. 
Fortunately, it is clear that to obtain the $E_3$-page table defined as [see Eq. \ref{eqn:Epages}]
\begin{equation}
E_{3}^{p,-n} := \text{Ker}(d_{2}^{p,-n}) / \text{Im}(d_{2}^{p-2,-(n-2)}), 
\end{equation} 
the only possibly non-trivial quantity that we need to compute is the kernel of $d_2^{0,-3}:E_2^{0,-3}\rightarrow E_2^{1,-4}$. 
This differential $d_2^{0,-3}$ in fact has a clear physical meaning under the topological phenomena interpretation, which is the process of symmetry-allowed band inversions or chemical potential shifts at high-symmetry points that generate gapless points in 2-cells. 
In the following we will adapt this interpretation to show that $d_2^{0,-3}$ is in fact a trivial mapping. 

To this end, we write down a low-energy effective BdG model that describes the process of $d_2^{0,-3}$
\begin{eqnarray}
H_{d_2^{0,-3}} &=& ( \boldsymbol{k}^{2} - \mu) \rho_{z} \otimes \sigma_{0} \otimes \tau_{z} + v_{x} k_{x} \rho_{x} \otimes \sigma_{z} \otimes \tau_{0}  \nonumber
\\
&+& v_{y} k_{y} \rho_{y} \otimes \sigma_{0} \otimes \tau_{z},  
\label{eqn:bandinvH}
\end{eqnarray}
where $\rho_i$, $\sigma_i$, and $\tau_i$ are Pauli matrices for orbitals with even- and odd-parity, the two spin species, and  particle and hole. This model is invariant under the time-reversal operation $\Theta = i \sigma_{y} \mathcal{K}$, the particle-hole operation $\Xi = \tau_{x} \mathcal{K}$, and the inversion $\mathcal{I} = \rho_{z} \tau_{z}$, where all three symmetry operations take momentum $\boldsymbol{k} \rightarrow -\boldsymbol{k}$.
When the chemical potential $\mu<0$, there is an odd parity occupied state at $\Gamma$ point. When  we tune the chemical potential $\mu>0$, the bands near the zero energy will undergo a band inversion at $\Gamma$ such that the occupied state at $\Gamma$ is now parity-even and create a pair of Dirac points in two opposite 2-cells in a direction perpendicular to the direction of $\boldsymbol{v} = (v_{x},v_{y})$. 
However, these two Dirac points are not protected and can be gapped out by a symmetry-allowed  $p$-wave pairing term 
\begin{equation}
H_{\Delta} = \Delta k_{x}  \rho_{0} \otimes \sigma_{z} \otimes \tau_{x} + \Delta k_{y} \rho_{z} \otimes \sigma_{0} \otimes \tau_{y}.
\end{equation}
Since this process of a band inversion at a 0-cell does not lead to gapless points in 2-cells, we find that the second differential $d_{2}^{0,-3}$ is a trivial mapping, which means $\text{Ker}(d_{2}^{0,-3})=E_{2}^{0,-3}$. 
We therefore conclude that the $E_3$-page is exactly the same as the $E_2$-page, which indicates the convergence of the spectral sequence. The table of the limiting page is therefore given by $E_{\infty}^{p,-n}=E_{2}^{p,-n}$ for $p=0,1,2$ and $n=3,4,5$ [see Table \ref{table:E2}]. 

Finally, equipped with the diagonal entries in the limiting-page table, we are now ready to calculate the full twisted equivariant K group $^{\phi}K^{\tau,-3}_{G}(T^2)$ for our system. 
We perform the calculation iteratively following the relations in Eq. \ref{eqn:sequence1} and Eq. \ref{eqn:sequence2}: 
The entries in the limiting page table that enter the iterative calculation are in fact the diagonal entries $E_{\infty}^{p,-(3+p)}$ with $p=0,1,2$ only. 
From Table \ref{table:E2} we find that both the classifications on the high-symmetry lines (1-cells) and on the 2D Brillouin zone (2-cell) are trivial, i.e. $E_{\infty}^{1,-4} = E_{\infty}^{2,-5}=0$ [see Table \ref{table:E2}]. 
The full K group is therefore directly given by the limiting page on the high-symmetry points (0-cells) $E_{\infty}^{0,-3}$. Specifically, we have  
\begin{equation}
^{\phi}K^{\tau-3}_{G}(T^{2}) \cong E_{\infty}^{0,-3} \cong E_{1}^{0,-3} \cong \z \times \z \times \z \times \z. 
\label{eqn:Kgroup_result}
\end{equation}

We emphasize two things about this result. First, the full classification we find in the momentum space by computing the twisted equivariant K group in Eq. \ref{eqn:Kgroup_result} is consistent with the full classification $\cal{C}$$(G)$ in Eq. \ref{eqn:tsc-classification} we find in the real space using the topological crystal scheme. This serves as a nice crosscheck for both of the classification schemes we use.  
Second, we find that the full twisted equivariant K group is solely determined by the classification on the high-symmetry points. Thus for any symmetry indicators we write down for our 2D superconductors, the values it takes \textit{should depend only on the symmetry information on the four high-symmetry points, and no information from the rest of the Brillouin zone}. Because of this finding from our K-group analysis, in the next section we will derive a set of symmetry indicators for our system that contains a known $0$D $\z$-invariant associated with a $\z$ in $E_{\infty}^{0,-3}$ on each of the high-symmetry points.

\section{Derivation for symmetry indicators}

In this section, we will give a faithful derivation of a set of symmetry indicators that serve as a diagnostics for the strong and weak phases classified in by the twisted equivariant K group in Eq. \ref{eqn:Kgroup_result}, where various subtleties in the derivation will be emphasized. Importantly, only indicators with such disentanglement in strong and weak phases can serve as faithful diagnostics for real-space boundary features. Since the classification we obtained from the twisted K theory is $\z \times \z \times \z \times \z$ for 2D class-DIII odd-parity superconductors with $\bf{p_2}$ symmetries, the corresponding indicators should be a linear combination of the 0D $\z$-invariants on the four TRIMs. 

In subsection A, we will discuss how to write down this 0D invariant on a given TRIM. Then in subsection B, we will show how to obtain the correct linear combinations for the 2D indicators of interest. 

\subsection{0D topological invariant}
We can deduce the form of a 0D topological invariant from the relations among the Karoubi's triples for 0D class-DIII BdG Hamiltonians with odd-parity pairing. The full symmetry group $G$ for these 0D systems are given in Eq. \ref{eqn:g}.  
In particular, since these 0D systems live on TRIMs, which are inversion-invariant points, the inversion symmetry $\mathcal{I}$ becomes an internal $\z_2$ symmetry. 
 
We start from examining the inverse elements of a Karoubi's triple $[(E, G), H, H_0]$ in the K group $^{\phi}K^{\tau-3}_{G}(T^{2})$. 
Importantly, due to the presence of the inversion symmetry, there exist more inverse elements besides $[(E, G), H_0, H]$.
To identify these additional inverse triples, we need to first identify the identity element in the K group. In the following, we consider two-band systems for simplicity.   
In such a case, both of the following Karoubi's triples 
\begin{eqnarray}
b_{0}^{+} &=& \left[ (E,I = \tau_{3}), H=\tau_{3}, H_{0} = \tau_{3}  \right],
\label{eqn:e0+}
\\
b_{0}^{-} &=& \left[ (E,I = -\tau_{3}), H=\tau_{3}, H_{0} = \tau_{3}  \right],
\label{eqn:e0-}
\end{eqnarray}
and their linear combinations belong to the identity class of the K group, where we have suppressed the labeling for the time-reversal and particle-hole symmetries in the symmetry group $G$. 
Here, the $H_0$'s in $b_{0}^{+}$ and $b_{0}^{-}$ correspond to the vacuum BdG Hamiltonians (made by vacuum normal states) whose ground states are even and odd under inversion, respectively, and the $H$'s in both triples are considered as trivial Hamiltonians. With the identity specified, we can now discuss the inverse element. 

For a given generator of the K group 
\begin{equation}
b_{1}^{+} = \left[ (E,I = \tau_{3}), H=-\tau_{3}, H_{0} = \tau_{3}  \right]
\end{equation}
with $H$ being a BdG Hamiltonian with an even-parity ground state, we find that the triple whose $H$ has an odd-parity ground state 
\begin{equation}
b_{1}^{-} = \left[ (E,I = -\tau_{3}), H=-\tau_{3}, H_{0} = \tau_{3}  \right]  
\end{equation}
is in fact an inverse element. 
In other words, the triple 
\begin{align}
\tilde{b}&=b_{1}^{+} + b_{1}^{-}\nonumber\\
&= \left[ (E,I = \tau_{3} \oplus -\tau_{3}), H=-\tau_{3} \oplus -\tau_{3}, H_{0} = \tau_{3} \oplus \tau_{3}  \right]
\end{align}
is an identity triple. 
To show the equivalence between $H$ and $H_0$ in $\tilde{b}$, we explicitly construct an adiabatic path that connects the $H$ and $H_0$ without closing the gap. 
The path we find has the following form 
\begin{equation}
H_{t} = t H_{0} + (1-t) H + H_{\Delta}(t),
\end{equation}
where 
\begin{equation}
H_{\Delta}=
\Delta(t) \left( \begin{array}{cccc}
0 & 0 & 0 & 1
\\
0 & 0 & 1 & 0
\\
0 & 1 & 0 & 0
\\
1 & 0 & 0 & 0
\end{array} \right) 
\end{equation}
and $\Delta(t)$ is a smooth function such that $\Delta(0)=\Delta(1)=0$. The path $H_{t}$ is gapped everywhere while preserving the symmetry. We have thus shown that $b_{1}^{-}$ is the inverse element of $b_{1}^{+}$. 
Physically speaking, this implies that the numbers of the even-parity and odd-parity occupied states in the 0D BdG Hamiltonian of our interest are in fact related. Therefore, somewhat surprisingly, the 0D invariant in a way should depend only on the number of even-parity or the odd-parity occupied states, but not both.

Based on this information, we now write down an explicit expression for this 0D invariant. Such an invariant should be an integer that takes the triples in the corresponding K group $K\cong \z$ as inputs, which is therefore the image of a group homomorphism $\hat{f}: K \rightarrow \z$. 
Moreover, 
a valid form for $\hat{f}$ should satisfy the following constraints:  
The identity triples should lead to an output of $0$, and the inverse triple for a triple with output $\tilde{n}$ should lead to an output of $-\tilde{n}$. 
For instance, we require $\hat{f}(b_{0}^{+}) = \hat{f}(b_{0}^{-}) = 0$, and if $\hat{f}(b_{1}^{+}) =1$, we require $\hat{f}(b_{1}^{-}) = -1$. According to these constraints as well as the relation between the Hamiltonians with even- and odd-parity occupied states, it is not difficult to see that 
\begin{equation}
n^k=\frac{1}{2}( N^{+}[H(k)] - N^{+}[H_{0}(k)] )  
\label{eqn:0d_invariant}
\end{equation}
serves as a valid 0D invariant at a given TRIM $k$ for any triple $\left[ (E,G^k), H(k), H_{0}(k) \right]$, where $N^+[h(k)]$ is the number of even-parity occupied states of Hamiltonian $h$ at high-symmetry point $k$, and the coefficient $1/2$ accounts for the Kramers degeneracy. 
Physically speaking, this $\z$ invariant $n^k$, which corresponds to the classification on each high-symmetry point $k$, counts the number difference between the even-parity occupied states of the considered BdG Hamiltonian and the universal reference Hamiltonian $H_0$ defined in Eq. \ref{eqn:vacuum}. 
Through explicit calculations, one can verify that this topological invariant $n^k$ indeed is the homomorphism $\hat{f}: K \rightarrow \z$. 
This result was also obtained in Ref.~\onlinecite{ShiozakiIndicator}, but here we provide a more detailed derivation. 

Finally, we will review the expression of $n^k$ in the weak-pairing limit discussed in Ref. \onlinecite{ShiozakiIndicator}, which could be useful for the purpose of material search. 
This is because in the weak-pairing limit, where the odd-parity order parameter $\Delta$ is weak enough that it does not alter the normal-state parity data but just requires the particle-hole related BdG bands to have opposite parities, the indicators can be written solely in terms of the normal-state information and are readily applicable for \textit{ab initio} calculations. 

In the weak-pairing limit, the BdG Hamiltonian $H(k)$ in the 0D invariant $n^k$ in Eq. \ref{eqn:0d_invariant} has the form
\begin{equation}
H = 
 \left( \begin{array}{cc}
h & \Delta 
\\
\Delta^{\dagger} & -h^{*}
\end{array} \right), 
\end{equation}
where $h$ is the normal-state Hamiltonian, and we take the odd-parity order parameter $\Delta\rightarrow 0$. We then write down the vacuum BdG Hamiltonian $H_{0}$ following our discussion in Eq. \ref{eqn:vacuum} as 
\begin{equation}
H_{0} = 
 \left( \begin{array}{cc}
\mathds{1} & 0 
\\
0 & -\mathds{1}
\end{array} \right),
\label{eq:Href}
\end{equation}
where $H_{0}$ is in the same basis as $H$. Since the parity data of these BdG Hamiltonians are related to those of the normal Hamiltonian as 
\begin{align}
&N^{+}[H] = N^{+}[h] + N^{+}[-h^{*}]\nonumber\\
&N^{+}[H_{0}] = N^{+}[\mathds{1} ] + N^{+}[-\mathds{1}], 
\end{align}
we obtain the following weak-pairing expression of Eq. \ref{eqn:0d_invariant}
\begin{equation}
n = \left( N^{+}[h] + N^{+}[-h^{*}] \right) -  \left( N^{+}[\mathds{1}] + N^{+}[-\mathds{1}] \right).
\label{eqn:n_mid}
\end{equation}
Since the particle-hole and the inversion symmetries anticommute [see Eq. \ref{eqn:g}], we can further simplify the expression using the relations
\begin{eqnarray}
N^{+}[-h^{*}] &=& \bar{N}^{-}[h],
\label{eqn:Nrelation1}
\\
N^{+}[-\mathds{1}] &=& \bar{N}^{-}[\mathds{1}], 
\label{eqn:Nrelation2}
\end{eqnarray}
where $\bar{N}^{-}[H]$ denotes the number of odd-parity unoccupied state of a BdG Hamiltonian $H$.
Furthermore, by definition we also have 
\begin{equation}
N^{-}[h] + \bar{N}^{-}[h] = \bar{N}^{-}[\mathds{1}]. 
\label{eqn:Nrelation3}
\end{equation}
With the relations in Eq. \ref{eqn:Nrelation1}-\ref{eqn:Nrelation3}, we can rewrite Eq. \ref{eqn:n_mid} and arrive at the final weak-pairing expression for the 0D invariant at high-symmetry point $k$ 
\begin{equation}
n^k = \frac{1}{2}( N^{+}[h(k)] - N^{-}[h(k)] ).
\label{eqn:n_weak}
\end{equation}
This weak-pairing expression of $n^k$ depends only on the normal-state parity data $N^{\pm}[h(k)]$. 

\subsection{The symmetry-indicator group for 2D inversion-symmetric superconductors}
Equipped with the 0D invariants at TRIMs, in this subsection we will proceed to derive a set of symmetry indicators for the 2D time-reversal and odd-parity superconductors with $\bf{p_2}$ symmetries.  
Such indicators should depend on accessible information of the systems, and are capable of diagnosing not just the bulk topology, but also the boundary modes. 
In fact, the type of boundary modes of a given superconducting state would be readily known if we knew to which class of topological crystal the superconductor corresponds.  
For instance, as we discussed in Section II, a topological superconductor constructed by one {$d_b=2$ building block exhibits helical Majorana edge modes, whereas that constructed by two $d_b=2$ blocks is a higher-order phase with two inversion-related Majorana Kramers pairs on the boundary. Moreover, various $d_b=1$ block configurations that has no blocks passing through the inversion center can also lead to weak phases with Majorana bands on some given set of edges.  
As for the superconducting states built by $d_b=0$ blocks, these atomic superconductors do not carry protected boundary modes, and are considered as topologically trivial. 

The information of block construction, however, is typically unknown for a given superconductor of interest. Instead, what is often more accessible is the symmetry data in the momentum space, namely the symmetry eigenvalues of occupied BdG bands\footnote{Symmetry data of occupied normal bands at TRIMs are sufficient in the weak-pairing regime.}. 
In particular, for the 2D superconductors of our interest, we have shown in Eq. \ref{eqn:Kgroup_result} that the bulk classification, and thus the boundary features, can be solely determined by the parity data of the occupied BdG bands at TRIMs. The task of deriving boundary-diagnosing symmetry indicators therefore boils down to identifying the linear combinations of the 0D invariants at TRIMs that are associated with each class of real-space block construction. In particular, since the ASC are considered trivial, we will identify and remove them from the states we derive indicators for. 

Mathematically speaking, the group of the 0D-invariants for our case of 2D superconductors preserving symmetry group $G$ is given by the limiting page $E_{\infty}^{0,-3}\cong\z \times \z \times \z \times \z$, whereas that from the ASC is given by the image of a homomorphism 
\begin{align}
f_0: \mathcal{C}_{0}(G) \rightarrow E_{\infty}^{0,-3}.  
\label{eqn:f0}  
\end{align}
As we defined in Section II, $\mathcal{C}_{0}(G)$ is the Abelian group of the atomic superconductors. We can therefore span $\mathcal{C}_{0}(G)$ by the real-space generators $\{{\bf e}_{{\bf r}}\}$, which are the ASC built by placing $d_b=0$ building blocks at the Wyckoff position ${\bf r}$ in all unit cells.  
Since the homomorphism in Eq.~\ref{eqn:f0} gives the 0D invariants at the four TRIMs for ASC, to quotient out the 0D-invariant combinations of ASC from those of all symmetry-preserving 2D superconductors, we need to compute the quotient group $E_{\infty}^{0,-3}/$Im($f_0$). 

To this end, we span the limiting page $E_{\infty}^{0,-3}$ and the group of 0D invariants Im($f_0$) for ASC in chosen sets of basis.  
When working with specific representation of $E_{\infty}^{0,-3}$, we find it more convenient to work with ``addition" instead of ``multiplication" for the group operation, and will switch to the former in the rest of the paper.

For the limiting page $E_{\infty}^{0,-3}$, it is natural to span it as 
\begin{align}
E_{\infty}^{0,-3}=\z[{\bf b}_{\Gamma}]+\z[{\bf b}_{X}]+\z[{\bf b}_{Y}]+\z[{\bf b}_{M}],
\label{eq:spanlimiting} 
\end{align}
where $\{{\bf b}_{k}\}$ are the momentum-space generators of the 0D invariants at TRIMs. Similarly, the group of ASC $\mathcal{C}_{0}(G)$ can be spanned as
\begin{align}
\mathcal{C}_{0}(G)=\z[{\bf e}_{(0,0)}]+\z[{\bf e}_{(\frac{1}{2},0)}]+\z[{\bf e}_{(0,\frac{1}{2})}]+\z[{\bf e}_{(\frac{1}{2},\frac{1}{2})}],
\label{eq:spanlimiting} 
\end{align}
where $\{{\bf e}_{\textbf{r}}\}$ are the generators of the real-space 0D invariants at the special Wyckoff positions ${\textbf{r}}$ in a unit cell.
Using this basis, we can write down the matrix representation for the homomorphism $f_0$, which maps the real-space invariants of ASC to the corresponding momentum-space parity data, as 
\begin{align}
M_{f_{0}} = \left(\begin{array}{cccc}
\boldsymbol{a}_{1} & \boldsymbol{a}_{2} & \boldsymbol{a}_{3} & \boldsymbol{a}_{4} \\
\end{array}\right),
\label{eq:Mf0}  
\end{align}
where each column vector $\boldsymbol{a}_{i}$ contains the set of 0D invariants at TRIMs generated by a ASC associated with a certain Wyckoff position in the real space.
In our case, we find
\begin{align}
M_{f_{0}} = \left(\begin{array}{cccc}
1 & 1 & 1 & 1 \\
1 & -1 & 1 & -1 \\
1 & 1 & -1 & -1 \\
1 & -1 & -1 & 1 \\
\end{array}\right),   
\label{eq:Mf0}  
\end{align}
where the bases are given by $f_0[\boldsymbol{a}_1,\boldsymbol{a}_2,\boldsymbol{a}_3,\boldsymbol{a}_4]=[\boldsymbol{b}_\Gamma,\boldsymbol{b}_X,\boldsymbol{b}_Y,\boldsymbol{b}_M]M_{f_0}$.

Although the group of the ASC 0D invariants Im$(f_{0})$ can be spanned by $\{\boldsymbol{a}_{i}\}$, these basis vectors $\{\boldsymbol{a}_{i}\}$ are in general not linearly independent. To obtain a set of linearly independent bases for Im$(f_{0})$, we compute the Smith normal form for the matrix $M_{f_{0}}$ by
\begin{equation}
UM_{f_{0}}V= \lambda 
\label{eq:snf}
\end{equation}
, where $U$ and $V$ are the transformation matrices for the momentum-space and real-space bases respectively, and
\begin{align}
\lambda = \left(\begin{array}{cccc}
1 & 0 & 0 & 0  \\
0 & 2 & 0 & 0  \\
0 & 0 & 2 & 0 \\
0 & 0 & 0 & 4 \\
\end{array}\right).
\label{eq:lambda}
\end{align}
From the fact that $M_{f_{0}}V = U^{-1} \lambda$, we can now extract the linearly independent basis vectors and span both $E_{\infty}^{0,-3}$ and Im$(f_{0})$ in the new bases. 
Specifically, the new real-space basis vectors are given by 
\begin{align}
M_{f_{0}}V = \left(\begin{array}{cccc}
\boldsymbol{a}'_{1} & \boldsymbol{a}'_{2} & \boldsymbol{a}'_{3} & \boldsymbol{a}'_{4} \\
\end{array}\right), 
\label{eq:Mf0new} 
\end{align} 
where $\{\boldsymbol{a}'_{i}\}$ are column vectors rotated by the transformation matrix $V$ from $\{\boldsymbol{a}_{i}\}$,  
and the new momentum-space basis vectors are given by 
\begin{align}
U^{-1} = \left(\begin{array}{cccc}
\boldsymbol{b}'_{1} & \boldsymbol{b}'_{2} & \boldsymbol{b}'_{3} & \boldsymbol{b}'_{4} \\
\end{array}\right),   
\label{eq:Ubasis}  
\end{align}
where $\{\boldsymbol{b}'_{i}\}$ are column vectors rotated by $U^{-1}$ from $\{\boldsymbol{b}_{j}\}$ at $j$=TRIMs. 
Since the two sets of new bases are related by 
\begin{equation}
\boldsymbol{a}'_{i} = \boldsymbol{b}'_{i} \lambda_{i},
\end{equation}
where $\lambda_{i}$ denotes the diagonal element of $\lambda$, we can span the 0D invariant group for ASC and the limiting page  in the same set of linearly independent bases as  
\begin{align}
E_{\infty}^{0,-3}=\z[{\bf b}'_{1}]+\z[{\bf b}'_{2}]+\z[{\bf b}'_{3}]+\z[{\bf b}'_{4}],
\label{eq:spanlimiting_new} 
\end{align}
and 
\begin{align}
\text{Im}f_{0}=\z[{\bf b}'_{1}]+\z[2{\bf b}'_{2}]+\z[2{\bf b}'_{3}]+\z[4{\bf b}'_{4}].
\label{eq:spanlimiting_new} 
\end{align}
It is then straightforward to compute the group of symmetry indicators $\mathfrak{I}_{\text{SI}}$ by quotienting out the ASC 
\begin{align}
\mathfrak{I}_{\text{SI}} \equiv E_{\infty}^{0,-3}/{\text{Im}(f_{0})}=\z_2+\z_2+\z_4.
\label{eq:indicatorgp} 
\end{align}

Before we proceed to the explicit expressions for symmetry indicators, here we make three comments on the symmetry indicator group. We first make a brief comparison with existing literature\cite{Geier2020,Ono2020refined}. The group Im$(f_{0})$ is often referred to as ``$\{ \text{AI} \}$" in previous works, and the group of symmetry indicator is often defined as $\{ \text{BS} \}/\{ \text{AI} \}$, where $\{ \text{BS} \}$ is the set of 0D invariants at the high-symmetry momenta that satisfy certain set of compatibility relations. In our formalism, the elements in $E_{\infty}^{0,-3}$ are obtained by imposing the most complete compatibility relations such that every elements in $E_{\infty}^{0,-3}$ can be realized by some gapped Hamiltonians.

Secondly, we note that the indicator group $\mathfrak{I}_{\text{SI}}$ agrees with the group of topological superconductors $\mathcal{C}_\text{TSC}(G)$ obtained by the topological crystal approach [see Eq. \ref{eqn:real_indicators}]. This is expected since, in our case, the momentum-space  classification is entirely given by the classification of 0D topological invariants at TRIMs, i.e. $\mathcal{K}_{G}^{\tau-3}(T^{2}) \cong E_{\infty}^{0,-3}$. There is thus an isomorphism between the real-space classification $\mathcal{C}(G)$ and the classification of 0D topological invariants at TRIMs $E_{\infty}^{0,-3}$. Nonetheless, we point out that such an isomorphism does not exist when the full momentum-space classification also contains non-trivial contributions from 1D or 2D topological invariants defined on high-symmetry lines and at general points in Brillouin zone. In that case, the symmetry indicators can only provide partial information of the $\mathcal{C}_\text{TSC}(G)$. 

Finally, we emphasize that $\mathfrak{I}_{\text{SI}}$ and $\mathcal{C}_\text{TSC}(G)$ are only isomorphic to each other and there is no canonical mapping between them. The physical meanings of the symmetry indicators  $\mathfrak{I}_{\text{SI}}$ are therefore ambiguous unless we specify a mapping between $\mathfrak{I}_{\text{SI}}$ and $\mathcal{C}_\text{TSC}(G)$. In particular, only through such a mapping can the symmetry indicators have well-defined notions of strong and weak phases, and thus correspondence to boundary features [see discussion in Sec.~II].

\subsection{Explicit expressions for the symmetry indicators}
We now derive the explicit symmetry indicators that correspond to the symmetry indicator group $\mathfrak{I}_{\text{SI}}$. Importantly, we show how to disentangle the strong and weak phases and arrive at one purely strong $\z_4$ indicator and two purely weak $\z_2$ indicators. This and the next sub-sections are the key steps in our protocol.

From the Smith decomposition in Eq. \ref{eq:lambda}, it is clear that the transformation matrix $U$ contains independent momentum-space basis vectors that correspond to the $\z_4$ and $\z_2$ subgroups in the symmetry indicator group $\mathfrak{I}_{\text{SI}}$. 
Therefore, for a given superconducting state, it seems like we can simply obtain the explicit expression of its symmetry indicators by projecting its 0D invariants  
$\bar{{\bf{n}}}=[\bar{n}_{\Gamma},\bar{n}_{X},\bar{n}_{Y},\bar{n}_{M}]^{T}$ onto the new bases contained in $U$ as 
\begin{equation}
\boldsymbol{\nu} \equiv U \bar{{\bf{n}}}. 
\label{eqn:firstnu}
\end{equation}
Since the first column vector of $U$ is also a generator of Im$f_{0}$, the corresponding projection does not survive after we take the quotient. We will use $\boldsymbol{\nu}$ to denote the remaining  three components with this understanding. 

The indicators obtained in Eq. \ref{eqn:firstnu}, however, cannot serve as faithful boundary diagnostics.
This is because $\boldsymbol{\nu}$ obtained this way are \textit{not} guaranteed to correspond to purely strong or weak phases since there is in fact an important degree of freedom in the basis choice when obtaining the smith normal form in Eq.~\ref{eq:snf}. In particular, one can obtain a different momentum-space transformation matrix $\tilde{U}$ through basis transformations by unimodular matrices $L$ and $R$
\begin{eqnarray}
\lambda &=& U M_{f_{0}} V 
\\
&=& (U L^{-1}) (L M_{f_{0}} R^{-1}) (RV),
\label{eqn:LR}
\\
&=& \tilde{U} \tilde{M}_{f_{0}} \tilde{V}, 
\end{eqnarray}
while the symmetry indicator group $\mathfrak{I}_{\text{SI}}$ stays unchanged.  
Specifically, since we now have $\tilde{M}_{f_{0}} \tilde{V} =  \tilde{U}^{-1} \lambda$ under this new basis transformation, Im$f_{0}$ and $E_{\infty}^{0,-3}$ can be spanned respectively in new sets of linearly independent basis vectors  
\begin{align}
\tilde{M}_{f_{0}} \tilde{V}  &= \left(\begin{array}{cccc}
\tilde{\boldsymbol{a}}_{1} & \tilde{\boldsymbol{a}}_{2} & \tilde{\boldsymbol{a}}_{3} & \tilde{\boldsymbol{a}}_{4} \\
\end{array}\right),\\
\tilde{U}^{-1} &= \left(\begin{array}{cccc}
\tilde{\boldsymbol{b}}_{1} & \tilde{\boldsymbol{b}}_{2} & \tilde{\boldsymbol{b}}_{3} & \tilde{\boldsymbol{b}}_{4} \\
\end{array}\right). 
\label{eq:Mf0new}  
\end{align} 
Since these basis vectors again satisfy the same identification $\tilde{\boldsymbol{a}}_{i} = \tilde{\boldsymbol{b}}_{i} \lambda_{i}$, the indicator group $\mathfrak{I}_{\text{SI}}$ is unchanged. However, the explicit form of the symmetry indicators is now given by
\begin{equation}
\tilde{\boldsymbol{\nu}} \equiv \tilde{U} \bar{{\bf{n}}}.
\label{eqn:indicator_new}
\end{equation}
It is therefore clear that there is no unique explicit expression for indicators that correspond to the indicator group $\mathfrak{I}_{\text{SI}}$. Although all these different forms of symmetry indicators are equally valid for describing the bulk topology, only one of them written in the ``properly chosen'' bases has a simple decomposition between strong and weak phases, and therefore a transparent correspondence to the real-space boundary type. 

Here we identify the ``proper bases'' $L$ and $R$ by requiring the resulting symmetry indicators $\boldsymbol{\nu}$ to be consistent with our real-space topological invariants $\mathfrak{\Delta}$ defined in Eq. \ref{eq:rspace_inv}. The reason is because the real-space invariants $\mathfrak{\Delta}$, which are constructed based on the building-block constructions, 
naturally provide clear diagnosis for the boundary features [see Sec. II B]. To be explicit, we find the appropriate transformation matrices $L$ and $R$ in Eq.~\ref{eqn:LR} and arrive at the indicators $\tilde{\boldsymbol{\nu}}$ in Eq.~\ref{eqn:indicator_new} by demanding that the map from the real-space invariants to the symmetry indicators $\varphi:  \mathfrak{\Delta} \rightarrow \tilde{\boldsymbol{\nu}}$ takes the form
\begin{eqnarray}
\varphi &:& (\delta_{s},\delta_{\{\bar{1}|00\}}, \delta_{\{1|10\}}, \delta_{\{1|01\}}) \nonumber
\\
&\mapsto& (\delta_{s}+2\delta_{\{\bar{1}|00\}},  \delta_{\{1|10\}}, \delta_{\{1|01\}}), 
\label{eqn:phi}
\end{eqnarray}
where $\mathfrak{\Delta} = (\delta_{s},\delta_{{\bar{1}|00}}, \delta_{{1|10}}, \delta_{{1|01}})$ satisfies $(2,0,0,0) \cong (0,1,0,0)$. The map $\varphi$ defined this way ensures the separation between the strong and weak phases. 
The actual matrices we find are
\begin{align}
L = R = \left(\begin{array}{cccc}
0 & 0 & 0 & 1  \\
0 & 0 & -1 & 0  \\
0 & -1 & 0 & 0 \\
1 & 0 & 0 & 0 \\
\end{array}\right), 
\label{eq:LR-new}
\end{align}
where the corresponding momentum-space transformation matrix has the form 
\begin{align}
\tilde{U} = \left(\begin{array}{cccc}
0 & 0 & 0 & 1 \\
0 & 0 & 1 & 1 \\
0 & 1 & 0 & 1 \\
1 & 1 & 1 & 1 \\
\end{array}\right).
\label{eq:U-indicators}
\end{align}
Substituting Eq.~\ref{eq:U-indicators} into Eq.~\ref{eqn:indicator_new}, we arrive at the explicit expressions of the symmetry indicators 
\begin{align}
&\kappa=\sum_{j}\tilde{U}_{4j}\bar{n}_j=n_{\Gamma}+n_{X}+n_{Y}+n_{M}~~~{\text{mod}}~4\nonumber\\
&\nu_x=\sum_{j}\tilde{U}_{3j}\bar{n}_j=n_{X}+n_{M}~~~{\text{mod}}~2\nonumber\\
&\nu_y=\sum_{j}\tilde{U}_{2j}\bar{n}_j=n_{Y}+n_{M}~~~{\text{mod}}~2, 
\label{eq:indicators} 
\end{align} 
and we denote the set of symmetry indicators as $\boldsymbol{\nu}=(\kappa,\nu_x,\nu_y)$. 
In the rotated bases we choose, the generator of the $\z_4$ indicator $\kappa$ corresponds solely to the strong phases generated by the 2D building blocks, and the $\z_2$ indicators $\nu_x$ and $\nu_y$ correspond to the weak phases protected by translations in $x$- and $y$-direction generated by the 1D building blocks. We dubbed such a map a \textit{canonical map} and the corresponding basis \textit{canonical basis}. In the next section, we will discuss in details the basis-matching procedure between the real- and momentum-space approaches and show that Eq.~\ref{eq:indicators} are indeed the symmetry indicators in the the canonical basis.

Before we proceed to the next section, we give a final example to illustrate the importance to properly fix the basis ambiguity. Suppose we computed the Smith normal form for $M_{f_0}$ directly without identifying the canonical basis and arrived at indicators $(\kappa', \nu'_{x}, \nu'_{y})$.
For a state with real-space invariants $\mathfrak{\Delta}=(0,1,1,0)$, we would find the momentum-space indicators to be $(\kappa', \nu'_{x}, \nu'_{y}) = (0,1,0)$. 
Naively we would expect that $(\kappa', \nu'_{x}, \nu'_{y}) = (0,1,0)$ indicates a purely weak superconductor. However, this state is in fact a combination of a second-order strong phase and a weak phase according to its real-space invariants $\mathfrak{\Delta}$.  
It is therefore crucial to identify the canonical bases for the symmetry indicators to avoid confusions when applying the indicators to realistic models or real materials.

\subsection{Basis-matching procedure for the case study} 
Practically speaking, the canonical map in Eq.~\ref{eqn:phi} can be explicitly constructed by considering minimal models of the building blocks in the topological crystals. This ``basis-matching procedure'' between the momentum and the real spaces is our central result, which provides a general way to derive unambiguous symmetry indicators that diagnoses the boundary types.
In in subsection, we construct the map $\varphi: \mathfrak{\Delta} \rightarrow \boldsymbol{\nu}$ through explicit minimal models for the topological crystal states. The purpose of this subsection is to show Eq.~\ref{eqn:phi}, where the resulting symmetry indicators include one for purely strong phases ($\kappa$) and two for purely weak phases ($\nu_{x/y}$).

\begin{figure}
\includegraphics[width=1\columnwidth]{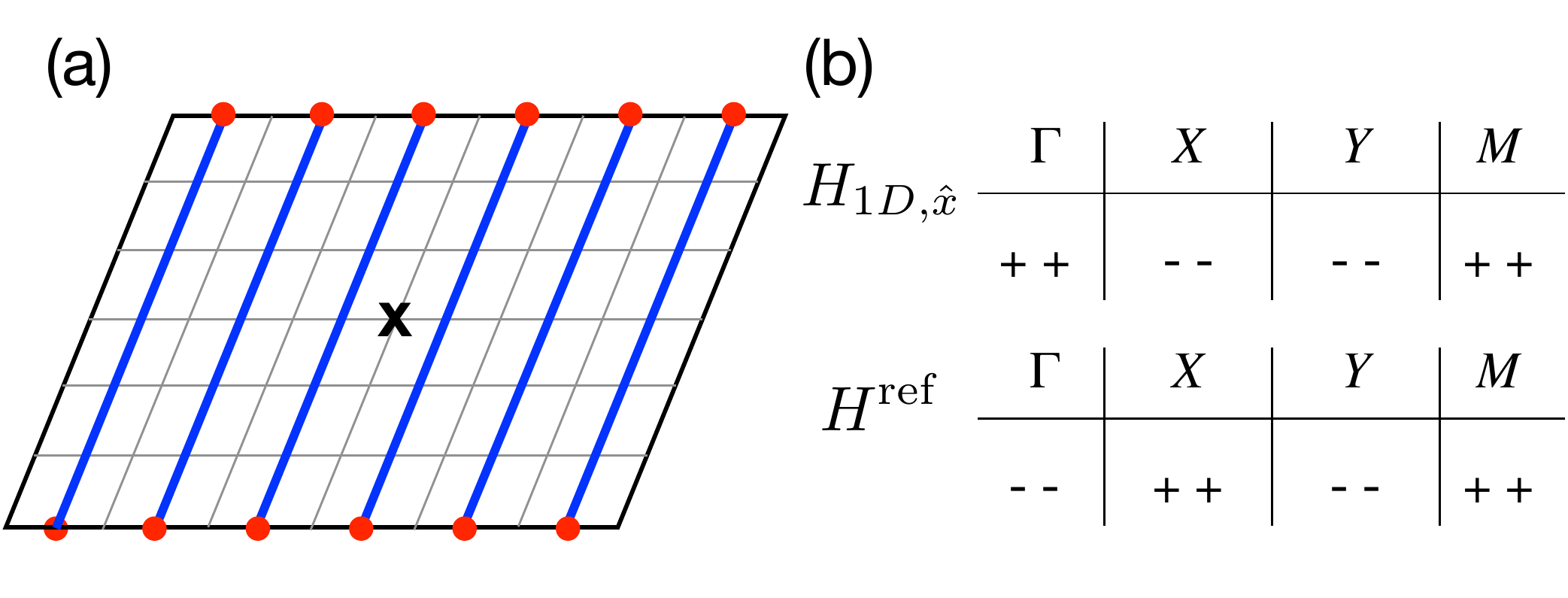}
\caption{(a) The building-block construction for the representative topological crystal we consider for the $\boldsymbol{\nu}=(0,1,0)$ weak phase. The grey squares are the unit cells, the blue lines are 1dTSCs, each red point represents a Majorana Kramers pair, and the ``x'' symbol marks the inversion center. (b) The parity data (of the occupied BdG bands) for the weak-phase model $H_{1D,\hat{x}}$ and the reference Hamiltonian $H^{\text{ref}}$. 
}
\label{fig:model_weak}
\end{figure}

\subsubsection{Weak phases}
We start from a $d_b=1$ topological crystal state with the real-space invariant $\mathfrak{\Delta}=(\delta_{s},\delta_{\{\bar{1}|00\}}, \delta_{\{1|10\}}, \delta_{\{1|01\}}) = (0,0,1,0)$. Such a state is a weak phase protected by the translational symmetry in $\hat{x}$ ($T_x$) alone. First, we construct a minimal model for this state 
\begin{align}
H_{1D,\hat{x}}({\bf{k}})=-(\cos k_y-\mu)\tau_zs_0+\sin k_y\tau_xs_z   
\label{eq:1dblock}
\end{align}
by stacking $\hat{y}$-directional $d_b=1$ building blocks in the $\hat{x}$ direction with negligible couplings between blocks. Here, $s_i$ and $\tau_i$ are Pauli matrices for the spin and the particle-hole spaces, respectively. Specifically, to obtain a phase that is protected solely by the translational symmetry $T_y$, we place one $d_b=1$ block per unit cell on the 1-cell $e^{(1)}_{3}$ in Fig. \ref{fig:real_cells}(a) [see Fig. \ref{fig:model_weak}(a)]. Second, we follow Eq. \ref{eqn:vacuum} and define the reference Hamiltonian as 
\begin{align}
H^{{\text{ref}}}({\bf{k}})=\tau_zs_0. 
\label{eq:Hrefweak}
\end{align}
Finally, we write down the inversion operator associated with such a configuration. The inversion operator $\mathcal{I}({\bf{k}})$ generally acquires momentum dependence when the unit cell containing the inversion center does not preserve the global inversion symmetry on its own, such as a system with sublattice whose inversion center is on a unit-cell corner. For the current configuration, since we assume the inversion center is at $(0,0)$ and the $d_b=1$ blocks are all away from the inversion center, the inversion operator has the momentum dependence of  
\begin{align}
\mathcal{I}_{1D,\hat{x}}({\bf{k}})=e^{ik_x}\tau_zs_0.     
\end{align} 

When we set $\mu>1$ and $\mu<1$, the model $H_{1D,\hat{x}}({\bf{k}})$ is well-known to be in the trivial and non-trivial weak phases, respectively. Given the parity data of $H_{1D,\hat{x}}({\bf{k}})$ and the reference Hamiltonian $H^{{\text{ref}}}$ calculated using the inversion operator $\mathcal{I}_{1D,\hat{x}}({\bf{k}})$ [see Fig. \ref{fig:model_weak}(b)], we find the indicators defined in Eq. \ref{eq:indicators} to be 
\begin{align}
&\boldsymbol{\nu}=(\kappa,\nu_x,\nu_y)=(0,0,0)~~~~~\text{for}~\mu>1,\nonumber\\
&\boldsymbol{\nu}=(\kappa,\nu_x,\nu_y)=(0,1,0)~~~~~\text{for}~\mu<1,    
\end{align}
respectively. Therefore, we find the map $\varphi$ defined in Eq. \ref{eqn:phi} to be $\varphi: (0,0,1,0) \mapsto (0,1,0)$. 

Similarly, we perform an analogous analysis for $T_y$-protected weak phases by constructing a model with the real-space invariant $\mathfrak{\Delta} = (0,0,0,1)$ following the discussion in Sec. II B 2. We find the indicators to be $(\kappa,\nu_x,\nu_y)=(0,0,0)$ and $(0,0,1)$ for the trivial phase ($\mu>1$) and the non-trivial weak phase ($\mu<1$), respectively. 
This again shows that the map $\varphi: (0,0,0,1) \mapsto (0,0,1)$, as expected. 
Given the Majorana end modes from each of the $d_b=1$ blocks, we expect that these weak phases protected by translations $T_x$ and $T_y$ support Majorana bands on edges along $\hat{x}$ and $\hat{y}$, respectively. 

\begin{figure}
\includegraphics[width=1\columnwidth]{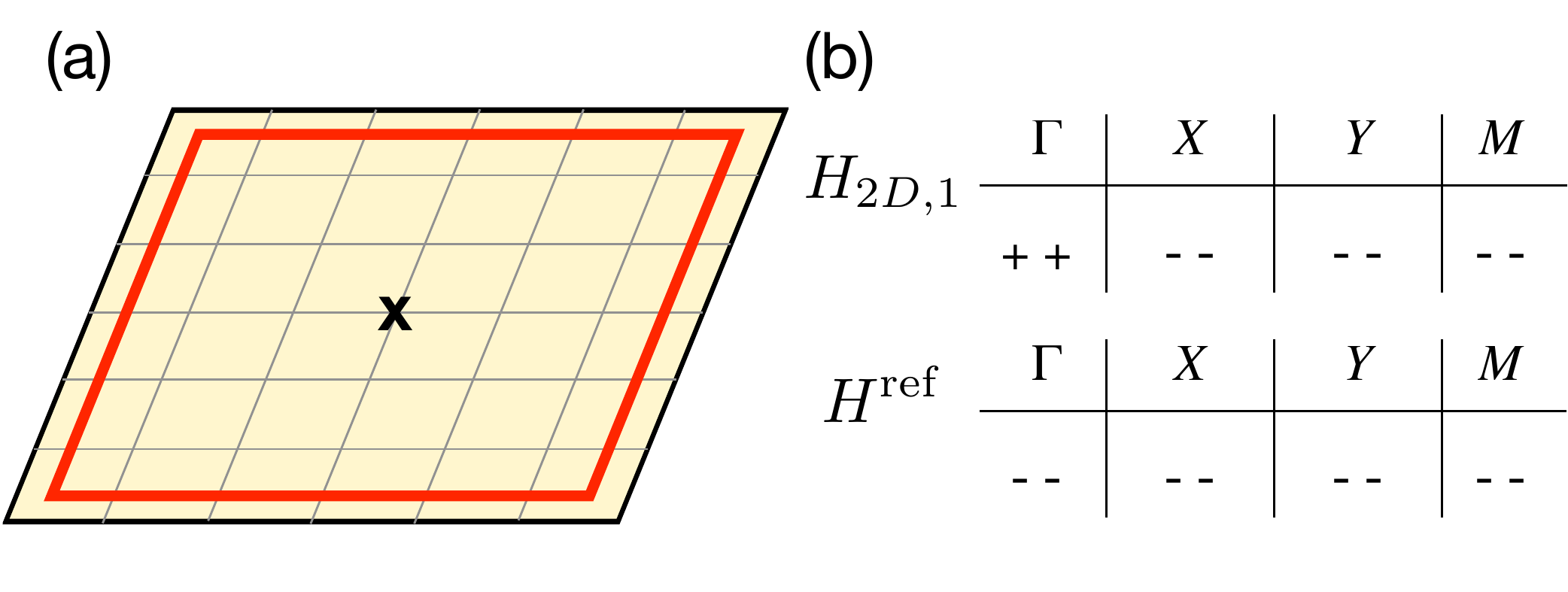}
\caption{(a) The building-block construction for the representative topological crystal we consider for the $\boldsymbol{\nu}=(1,0,0)$ strong phase. The grey squares are the unit cells, the yellow 2D object represents a 2dTSC, the red square represents a pair of counter-propagating Majorana edge modes, and the ``$\times$'' symbol marks the inversion center. (b) The parity data (of the occupied BdG bands) for the strong-phase model $H_{2D,1}$ and the reference Hamiltonian $H^{\text{ref}}$.
}
\label{fig:model_strong}
\end{figure}
\subsubsection{First-order strong phase}
Next, we move on to the strong topological superconductors with $\mathfrak{\Delta}= (1,0,0,0)$. This state is constructed by placing one $d_b=2$ building block per 2-cell, and supports helical Majorana edge modes [see Fig. \ref{fig:model_strong}(a)], and we consequently consider the model 
\begin{align}
H_{2D,1}({\bf{k}})=&-(\cos k_x+\cos k_y-\mu)\tau_zs_0\nonumber\\
&+\sin k_x\tau_xs_z+\sin k_y\tau_ys_0.   
\end{align}
For such a construction, the inversion operator is momentum independent 
\begin{align}
\mathcal{I}_{2D,1}({\bf{k}})=\tau_zs_0,    
\end{align}
and we can take the same reference Hamiltonian $H^{{\text{ref}}}({\bf{k}})$ as in Eq. \ref{eq:Href}. 
For $0<\mu<2$,
we have a non-trivial class-DIII 2D superconductor (that is compatible with the inversion symmetry), which is well-known to host helical Majorana edge modes. 
From the parity data of $H_{2D}({\bf{k}})$ at TRIMs [see Fig. \ref{fig:model_strong}(b)], we indeed find that the indicators are   
\begin{align}
&\boldsymbol{\nu}=(1,0,0).     
\end{align}
We have therefore established the map $\varphi: (1,0,0,0) \mapsto (1,0,0)$.

\begin{figure}
\includegraphics[width=1\columnwidth]{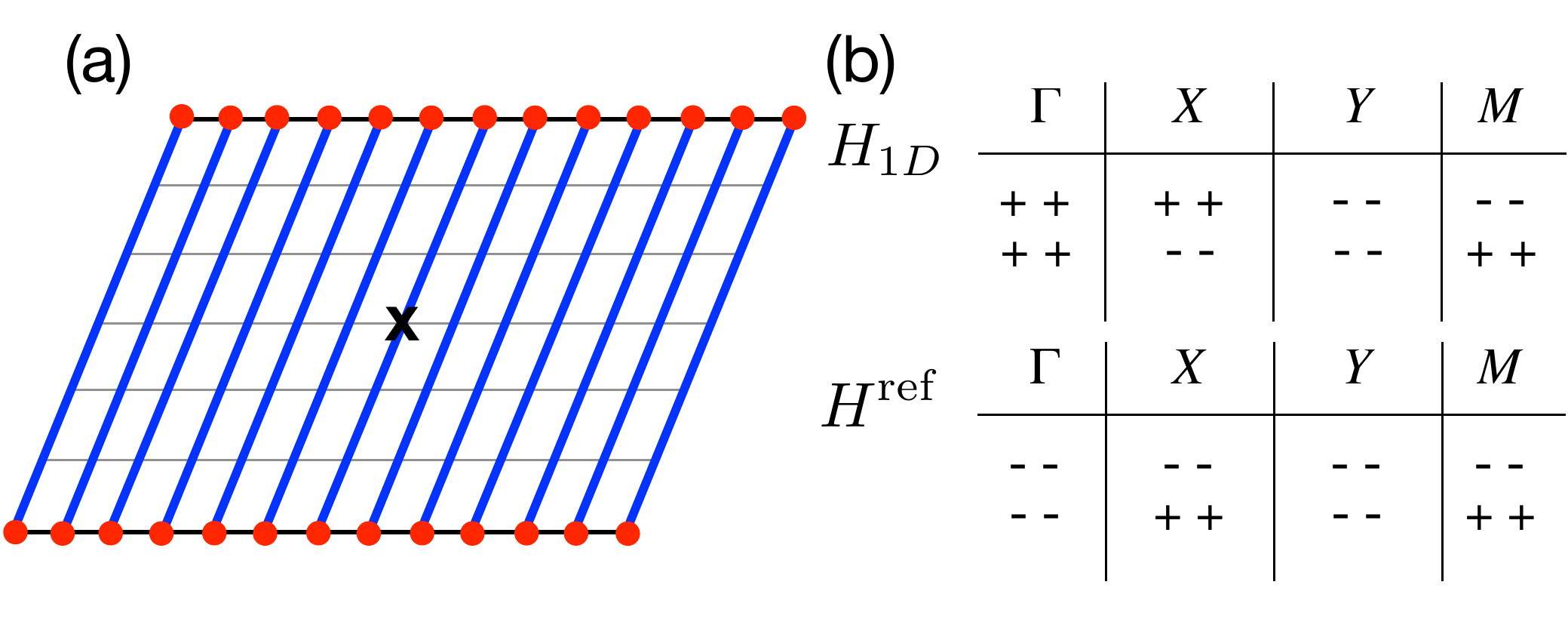}
\caption{(a) The building-block construction for the representative topological crystal we consider for the $\boldsymbol{\nu}=(2,0,0)$ higher-order strong phase. The grey squares are the unit cells, the blue lines represent 1dTSCs, each red point represents a Majorana Kramers pair, and the ``$\times$'' symbol marks the inversion center. (b) The parity data (of the occupied BdG bands) for the model $H_{1D}$ of the higher-order strong phase and for the reference Hamiltonian $H^{\text{ref}}$.
}
\label{fig:model_HO}
\end{figure}
\subsubsection{Second-order strong phase}
Finally, we consider the $d_b=1$ topological crystal state characterized by $\mathfrak{\Delta} = (0,1,0,0)\cong (2,0,0,0)$. Such a higher-order strong phase supporting inversion-protected corner Majoranas can be generated by placing either two $d_b=2$ building blocks per 2-cell, or two $d_b=1$ building blocks per unit cell, one across the inversion center ${\bf{r}}_0$ and the other one across the Wyckoff positions away from ${\bf{r}}_0$ [see Fig. \ref{fig:model_HO}(a)]. We choose to construct a minimal model using the $d_b=1$-block construction. Following the discussion in Sec. II B 2, for a geometry where the inversion center is at $(0,0)$, we place two $d_b=1$ blocks per unit cell, one at 1-cell $e^{(1)}_2$ and one at $e^{(1)}_3$ [see Fig. \ref{fig:model_HO}(a)]. 
The resulting model has the form  
\begin{align}
&H_{1D}({\bf{k}})=H_{0,0}({\bf{k}})\oplus H_{1/2,0}({\bf{k}}),\nonumber\\
&H_{0,0}({\bf{k}})=H_{1/2,0}({\bf{k}})=H_{1D,\hat{x}}({\bf{k}}), 
\label{eq:modelcornerMajo}
\end{align}
where $H_{0,0}$ and $H_{1/2,0}$ correspond to the two sets of $\hat{x}$-directional $d_b=1$ blocks that have negligible coupling to each other and are located at 1-cells $e^{(1)}_2$ and $e^{(1)}_3$, respectively. 

Since $H_{0,0}$ and $H_{1/2,0}$ are constructed by $d_b=1$ blocks on different sublattices, they transform under inversion operators with different momentum dependences
\begin{align}
&\mathcal{I}_{0,0}({\bf{k}})=\tau_zs_0, ~~~ \mathcal{I}_{1/2,0}({\bf{k}})=e^{ik_y}\tau_zs_0, 
\end{align}
and the inversion operator for the full system is given by $\mathcal{I}_{1D}({\bf{k}})=\mathcal{I}_{0,0}({\bf{k}})~\oplus~\mathcal{I}_{1/2,0}({\bf{k}})$. 
The parity data of $H_{1/2,0}({\bf{k}})$ is the same as that of the weak phase protected by translation in $\hat{y}$ [see Fig. \ref{fig:model_weak}(b)], whereas the parity data of $H_{0,0}({\bf{k}})$ is listed in Fig. \ref{fig:model_HO}(b). 
Since the resulting indicators from these two sets of parity data are $\boldsymbol{\nu}_{1/2,0}=(0,1,0)$ and $\boldsymbol{\nu}_{0,0}=(2,1,0)$, respectively, we find the indicators for the full Hamiltonian to be 
\begin{align}
\boldsymbol{\nu}=(0,1,0)+(2,1,0)=(2,0,0).  
\end{align}
We have therefore shown the map $\varphi: (0,1,0,0) \mapsto (2,0,0)$. 
This set of indicators corresponds to the strong phase with two inversion-related 0D Majorana Kramers pairs on the boundary.   

After performing the basis-matching procedure for the above topological crystal states, which are the generators for the entire topological superconductor group $\mathcal{C}_{TSC}(G)$, we have shown that the map $\varphi$ indeed takes the form of Eq.~\ref{eqn:phi} and the rotated bases we chose in Eq. \ref{eq:Mbasis} are indeed the canonical bases. As a result, our $\z_4$ indicator $\kappa$ indeed corresponds to just the strong phases generated by $d_b=2$ blocks and our $\z_2$ indicators $\nu_{x/y}$ indeed correspond to just the weak phases generated by $d_b=1$ blocks (away from the inversion center). 
The set of symmetry indicators $\boldsymbol{\nu}=(\kappa,\nu_x,\nu_y)$ therefore successfully decouples the strong and weak indices and serve as an effective boundary diagnostics for 2D class-DIII odd-parity superconductors with $\bf{p_2}$ symmetries. 

\subsection{Symmetry indicators in the weak-pairing regime}
As a final remark, we point out that it is practically useful to consider the weak-pairing expressions of our indicators in Eq. \ref{eq:indicators} 
\begin{align}
&\kappa=n_{\Gamma}+n_{X}+n_{Y}+n_{M}~~~{\text{mod}}~4\nonumber\\
&\nu_x=n_{X}+n_{M}~~~{\text{mod}}~2\nonumber\\
&\nu_y=n_{Y}+n_{M}~~~{\text{mod}}~2,\nonumber\\   
\label{eq:weakpairind1} 
\end{align}
by inserting the weak-pairing expression for a single 0D invariant $n^k$ in Eq. \ref{eqn:n_weak}    
\begin{align}
n^k=\frac{1}{2}(N^{+}[h(k)] - N^{-}[h(k)]),~~~k=\Gamma,X,Y,M. 
\label{eq:weakpairind2} 
\end{align}
Unlike the parity data of the BdG bands, here the numbers of even- and odd-parity occupied normal bands $N^{\pm}[h(k)]$ are often accessible by first principle calculations. Our indicators in the weak-pairing limit could therefore be applicable to material search for strong and weak topological phases among 2D time-reversal odd-parity superconductors.

\section{Applications on lattice models}
In this section, we further test our indicators $\boldsymbol{\nu}=(\kappa,\nu_x,\nu_y)$ in Eq. \ref{eq:indicators} against various lattice models where the corresponding topological crystal constructions are not obvious, but the parity data and boundary modes are known through analytical or numerical studies.   
In the following, we consider four BdG models of 2D time-reversal odd-parity superconductors studied in Ref. \onlinecite{WTe2HOTsc} and \onlinecite{ModelRXZ}.   

\subsection{Superconducting quantum spin Hall states}
\begin{figure}
\includegraphics[width=1\columnwidth]{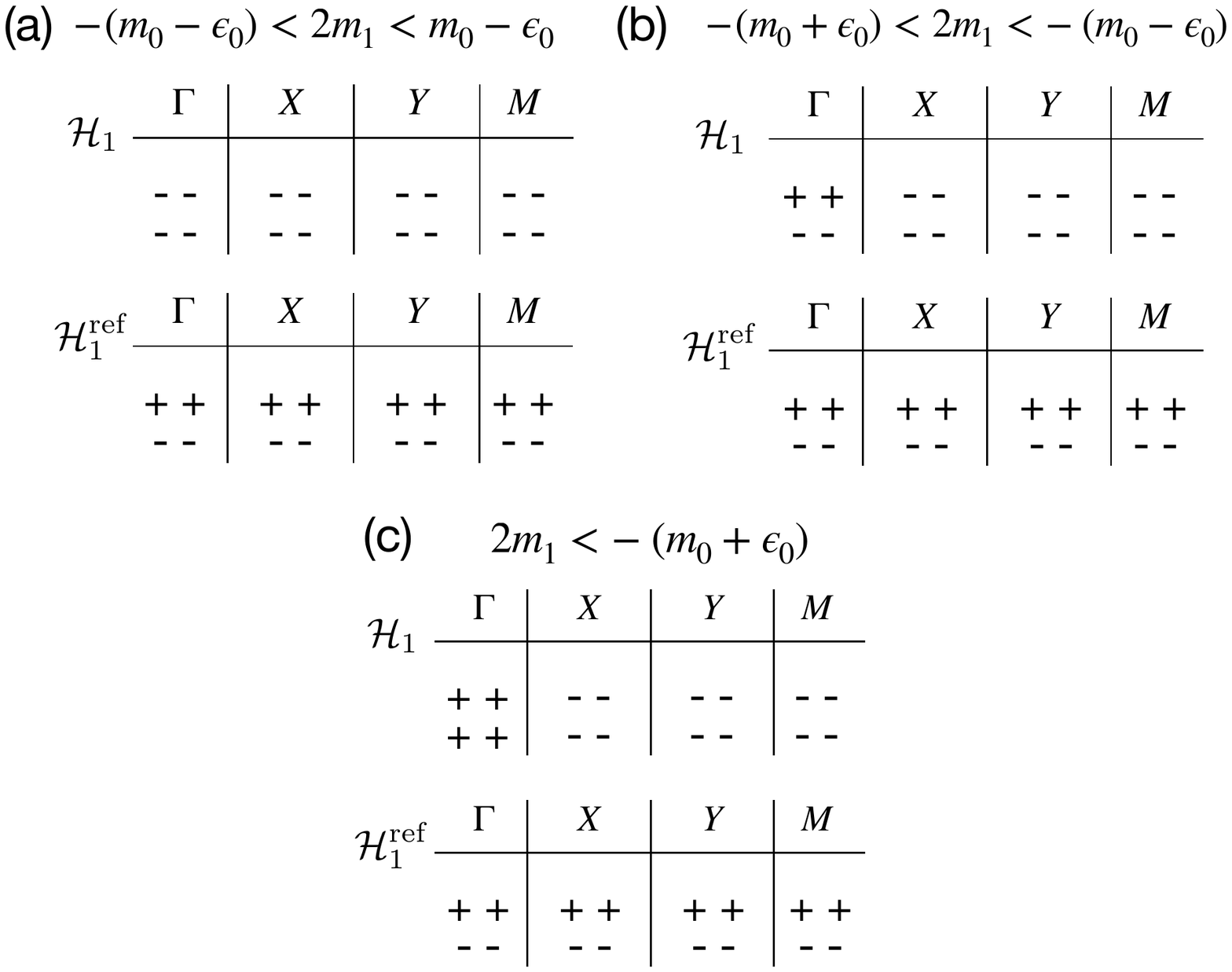}
\caption{The inversion eigenvalues of occupied BdG bands for the reference Hamiltonian $\mathcal{H}^{{\text{ref}}}_1$ and for the model $\mathcal{H}_1$ in (a) the trivial phase, (b) the topological phase with Majorana edge modes, and (c) the topological phase with Majorana corner modes. The parity data of $\mathcal{H}_1$ are quoted from Ref. \onlinecite{WTe2HOTsc}. 
}
\label{fig:egH1}
\end{figure}

We first study a minimal model proposed in Ref. \onlinecite{WTe2HOTsc} for superconducting phase transitions from a trivial phase to strong phases with Majorana edge modes and corner modes. 
Specifically, based on a general ``recipe'' for 2D inversion-protected higher-order superconductors proposed in the same paper,  
this model $\mathcal{H}_1=\sum_{\textbf{k}}\mathcal{H}(\textbf{k})$ 
\begin{align}
\mathcal{H}_1(\textbf{k}) &= \epsilon_0\hat{\tau}_z\otimes\hat{s}_0\otimes\hat{\rho}_0\nonumber\\
&+[m_0+m_1(\cos k_x+\cos k_y)]\hat{\tau}_z\otimes\hat{s}_0\otimes\hat{\rho}_z\nonumber\\
&+v\sin k_x\hat{\tau}_0\otimes\hat{s}_z\otimes\hat{\rho}_x+v\sin k_y\hat{\tau}_z\otimes\hat{s}_0\otimes\hat{\rho}_y\nonumber\\
&+\Delta\sin k_x\hat{\tau}_x\otimes\hat{s}_z\otimes\hat{\rho}_0+\Delta\sin k_y\hat{\tau}_y\otimes\hat{s}_0\otimes\hat{\rho}_z    
\label{eq:H1}
\end{align}
consists of a standard Bernevig-Hughes-Zhang like model for quantum-spin-Hall normal state and an odd-parity superconducting order parameter.  
Here $\hat{\tau}$, $\hat{s}$, and $\hat{\rho}$ are Pauli matrices for particle and hole, spin $s=\uparrow,\downarrow$, and orbital $\rho=s,p_-$. The Hamiltonian $\mathcal{H}_1$ obeys the time-reversal symmetry $\Theta_1=is_y\mathcal{K}$, $\textbf{k}\rightarrow-\textbf{k}$, the particle-hole symmetry $\Xi_1=\hat{\tau}_x\mathcal{K}$, $\textbf{k}\rightarrow-\textbf{k}$, and the inversion symmetry $\mathcal{I}_1$ $=\hat{\tau}_z\otimes\hat{\rho}_z$,  $\textbf{k}\rightarrow-\textbf{k}$, where the odd-parity pairing demands $\{\mathcal{I}_1,\Xi_1\}=0$.  

This lattice model $\mathcal{H}_1$ has been shown to exhibit two topological phase transitions in the parameter space of $\epsilon_0$, $m_0$, and $m_1$, one from a trivial phase to a phase with Majorana edge modes, and one from this phase with edge modes to a phase with Majorana corner modes. 
In particular, for a fixed $m_0$ and $\epsilon_0\geq 0$, it was shown\cite{WTe2HOTsc} analytically that 
\begin{align}
&\text{Trivial~boundary}:~~-(m_0-\epsilon_0)<2m_1<m_0-\epsilon_0\nonumber\\
&\text{Majorana edges}:~~~~-(m_0+\epsilon_0)<2m_1<-(m_0-\epsilon_0)\nonumber\\
&\text{Majorana corners}:~~~~~~~~~~~~~~~~~~~~~~2m_1<-(m_0+\epsilon_0),     
\label{eq:H1PT}
\end{align}
due to band inversions at $\Gamma$. 
Specifically, if we take the topologically trivial phase as a reference point, the BdG spectrum first undergoes a single band inversion to enter the phase with Majorana edges, then undergoes another band inversion to enter the phase with Majorana corners. 

To compute our indicators $\boldsymbol{\nu}=(\kappa,\nu_x,\nu_y)$ in Eq. \ref{eq:indicators} for these three topologically distinct phases, we first define the universal reference Hamiltonian as 
\begin{align}
\mathcal{H}^{\text{ref}}_1(\textbf{k}) &= \hat{\tau}_z\otimes\hat{s}_0\otimes\hat{\rho}_0  
\label{eq:H01}
\end{align}
following Eq. \ref{eqn:vacuum}. 
We then compute the parity data for the occupied BdG bands at TRIMs for $\mathcal{H}^{\text{ref}}_1$ and all three phases in $\mathcal{H}_1$ [see Fig. \ref{fig:egH1}]. The resulting indicators from these parity data are
\begin{align}
&(\kappa,\nu_x,\nu_y)=(4,0,0)~~~~{\text{for}}~~-(m_0-\epsilon_0)<2m_1<m_0-\epsilon_0\nonumber\\
&(\kappa,\nu_x,\nu_y)=(3,0,0)~~~~{\text{for}}~~-(m_0+\epsilon_0)<2m_1<-(m_0-\epsilon_0)\nonumber\\
&(\kappa,\nu_x,\nu_y)=(2,0,0)~~~~{\text{for}}~~2m_1<-(m_0+\epsilon_0)      
\label{eq:indicatorsH1}
\end{align}
for the three phases, respectively. The strong indicator $\kappa=4,3,2$ indicate phases with trivial boundaries, Majorana edge modes, and Majorana corner modes, respectively. The predicted boundary features by our indicators are therefore consistent with the boundary modes found analytically in Ref. \onlinecite{WTe2HOTsc} [see Eq. \ref{eq:H1PT}]. 

\subsection{Superconducting WTe$_2$ with odd-parity pairing}
\begin{figure}
\includegraphics[width=0.8\columnwidth]{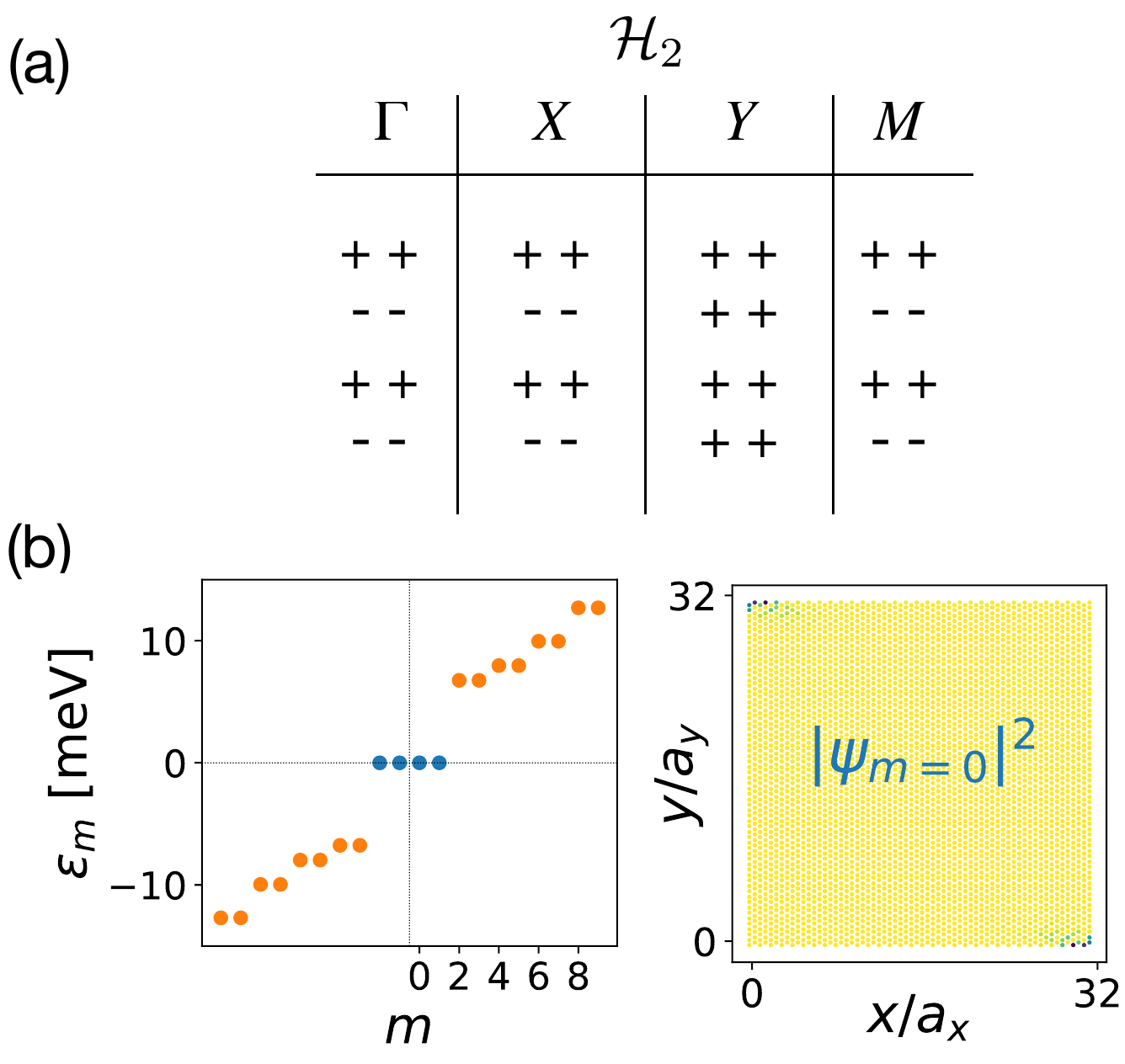}
\caption{(a) The inversion eigenvalues of occupied BdG bands for the reference Hamiltonian $\mathcal{H}^{{\text{ref}}}_2$ and for the model $\mathcal{H}_2$. (b) The numerical evidence for Majorana Kramers pairs on opposite corners in $\mathcal{H}_2$. The left panel shows the BdG spectrum against the eigenstate label $m$. There are four zero-energy eigenstates, which consist of two Kramers pairs localized on two opposite corners. The right panel shows the real-space profile $|\psi_{m=0}|^2$ of one of the zero-energy eigenstates on an open geometry, where $a_x$ and $a_y$ are lattice constants. The parity data and the figure in (b) are adapted from Ref. \onlinecite{WTe2HOTsc} with permission. 
}
\label{fig:egH2}
\end{figure}
The second model we consider is an 16-band BdG lattice model $\mathcal{H}_2=h_0+h_{\Delta}$ proposed in Ref. \onlinecite{WTe2HOTsc} for monolayer WTe$_2$, a superconducting quantum-spin-Hall material that preserves inversion symmetry. The normal part $h_0$ is an effective model based on \textit{ab initio} calculations for WTe$_2$\cite{WTe2_DFT_PRX,DFT_WTe2_SOC}. The pairing term $h_{\Delta}$ is the solution to linearized gap equations that belong to the time-reversal odd-parity $B_u$ irreducible representation in the point group $C_{2h}$\cite{WTe2HOTsc}.
This 2D superconducting state was numerically found to be a higher-order topological phase that hosts two Majorana Kramers pairs, one at each of the two opposite corners [see Fig. \ref{fig:egH2} (b)]. 

The parity data of this BdG model $\mathcal{H}_2$ was numerically calculated in Ref. \onlinecite{WTe2HOTsc} [see Fig. \ref{fig:egH2}(a)], which readily serve as the input for the indicator computation. 
Since the inversion operator has no momentum dependence due to the lattice structure of this material, the parity data of the reference Hamiltonian $H^{{\text{ref}}}_2={\text{diag}}[\mathds{1}_8,-\mathds{1}_8]$ are identical at the four high-symmetry points and do not affect the values of the indicators. 
The resulting indicators we find are therefore 
\begin{align}
&(\kappa,\nu_x,\nu_y)=(2,0,0), 
\label{eq:indicatorsWTe2}
\end{align}
which indicate a strong higher-order phase with inversion-protected Majorana Kramers pairs at opposite corners. The boundary type predicted by our indicators is therefore consistent with the boundary modes found numerically in Ref. \onlinecite{WTe2HOTsc} [see Fig. \ref{fig:egH2}(b)]. 

\subsection{First example with momentum-dependent inversion operator}
\begin{figure}
\includegraphics[width=1\columnwidth]{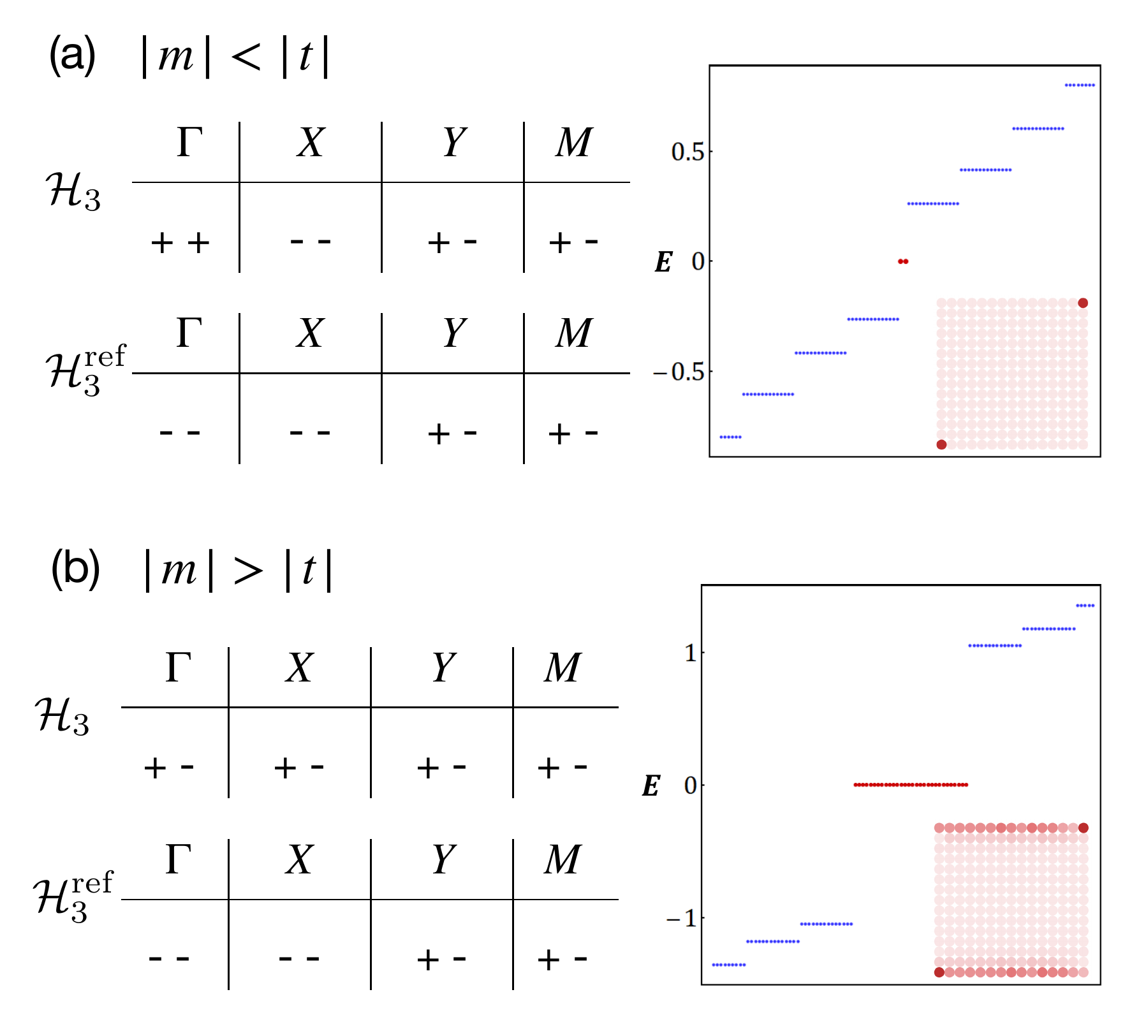}
\caption{The parity data and the numerical evidences for the boundary features for (a) $|m|<|t|$ and (b) $|m|>|t|$. 
The left panels show the inversion eigenvalues of occupied BdG bands for the model $\mathcal{H}_3$ and the reference Hamiltonian $\mathcal{H}^{\text{ref}}_3$. The right panels show the BdG spectrum $E$, where the red points correspond to the zero-energy eigenstates. The insets show the spatial profiles of one of the zero-energy eigenstates. The parity data of $\mathcal{H}_3$ and the figures on the right panels are adapted from Ref. \onlinecite{ModelRXZ} with permission.  
}
\label{fig:egH3}
\end{figure}
In the following, we study two systems with momentum-dependent inversion operators. This is a generic situation when a single unit cell alone is not inversion symmetric (with respect to the global inversion center $\textbf{r}_0$ or any of its integer multiple $m_1\textbf{r}_0$, $m_1\in\z$). For instance, sublattice systems whose inversion centers are on the unit-cell boundaries generally have momentum-dependent inversion operators.

We first study the following tight-binding BdG model proposed in Ref. \onlinecite{ModelRXZ}
\begin{align}
\mathcal{H}_3=&\sum_{{\bf{R}}}it\beta_{{\bf{R}},l}\alpha_{{\bf{R}}+a_x,l}+im_{\alpha}\alpha_{{\bf{R}},B}\alpha_{{\bf{R}}+a_y,A}\nonumber\\
&+im_{\beta}\beta_{{\bf{R}}+a_y,B}\beta_{{\bf{R}},A},      
\label{H1indicators}
\end{align}
which is written in terms of operators for the Majorana fermions $\alpha_{{\bf{R}},l}=(c_{{\bf{R}},l}+c^{\dagger}_{{\bf{R}},l})/\sqrt{2}$ and $\beta_{{\bf{R}},l}=(c_{{\bf{R}},l}-c^{\dagger}_{{\bf{R}},l})/(\sqrt{2}i)$ instead of those for the complex fermions $c^{\dagger}_{{\bf{R}},l}$ and $c_{{\bf{R}},l}$. Specifically, $\alpha_{{\bf{R}},l}$ and $\beta_{{\bf{R}},l}$ are operators for the two flavors of Majorana fermions on the same sublattice $l=A,B$ in the same unit cell centered at ${\bf{R}}$. 
Importantly, due to the sublattice structure of the model, $\mathcal{H}_3$ remains invariant under a momentum-dependent inversion operator 
\begin{align}
\mathcal{I}_3({\bf{k}})={\text{diag}}[1,e^{ik_y},-1,-e^{-ik_y}]      
\label{eq:P3}
\end{align}
in the basis of $[c_{{\bf{k}},A}$, $c_{{\bf{k}},B}$, $c^{\dagger}_{-{\bf{k}},A}$, $c^{\dagger}_{-{\bf{k}},B}]$, where $c_{{\bf{k}},l}$ and $c^{\dagger}_{{\bf{k}},l}$ are the Fourier transformed complex fermion operators. 

It was found analytically\cite{ModelRXZ} that $\mathcal{H}_3$ undergoes a phase transition at $|t|=|m|$ accompanied with a change in the boundary type. Specifically, both analytic and numerical results suggest that the $|m|<|t|$ phase exhibits Majorana corner modes [see Fig. \ref{fig:egH3}(a)], whereas the $|m|>|t|$ phase hosts additional 1D-like Majorana states on the $\hat{x}$-directional edges on top of Majorana corner modes [see Fig. \ref{fig:egH3}(b)]. 

We now compute our indicators for this model. Given that $\mathcal{H}_3$ is a four-band BdG Hamiltonian, we set the reference Hamiltonian to be 
\begin{align}
\mathcal{H}^{\text{ref}}_3({\bf{k}}) &= {\text{diag}}[\mathds{1}_2,-\mathds{1}_2].    
\label{eq:H03}
\end{align}
In Fig. \ref{fig:egH3} we display the parity data of the Fourier-transformed $\mathcal{H}_3$ computed in Ref. \onlinecite{ModelRXZ} and the parity data of $\mathcal{H}_3^{\text{ref}}$ we calculate using the momentum-dependent inversion operator $\mathcal{I}_3({\bf{k}})$.    
Since the model $\mathcal{H}_3$ has no time-reversal symmetry, we remove the $1/2$ factor in the indicators $\boldsymbol{\nu}$, which accounts for the Kramers degeneracy [see Eq. \ref{eq:indicators}]. Instead, we calculate the indicators $\tilde{\boldsymbol{\nu}}=(\tilde{\kappa},\tilde{\nu}_x,\tilde{\nu}_y)=2\boldsymbol{\nu}$ for time-reversal-broken superconductors. Specifically, we find    
\begin{align}
&(\tilde{\kappa},\tilde{\nu}_x,\tilde{\nu}_y)=(2,0,0)~~~~~{\text{for}}~~|t|>|m|\nonumber\\
&(\tilde{\kappa},\tilde{\nu}_x,\tilde{\nu}_y)=(2,1,0)~~~~~{\text{for}}~~|t|<|m|,    
\label{eq:indicatorsH3}
\end{align} 
which indicate an inversion-protected higher-order strong phase with two Majorana corner modes for the $|m|>|t|$ regime, and a combination of a weak phase on top of such a higher-order strong phase for the $|m|<|t|$ regime.  The boundary prediction made by our indicators is therefore consistent with the boundary types numerically found in Ref. \onlinecite{ModelRXZ} [see Fig. \ref{fig:egH3}(b)]. 

\subsection{Second example with momentum-dependent inversion operator}
\begin{figure}
\includegraphics[width=1\columnwidth]{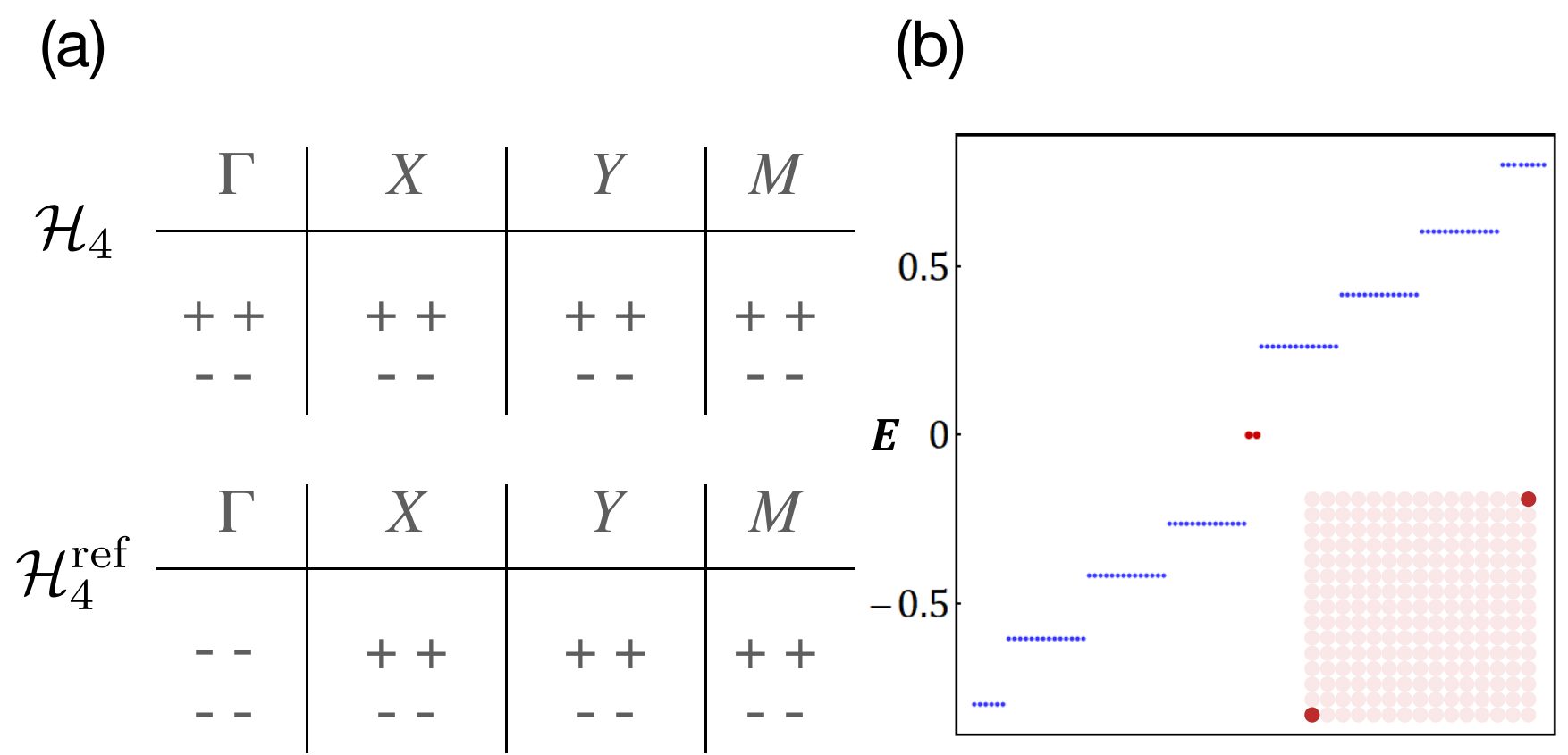}
\caption{(a) The inversion eigenvalues of occupied BdG bands for the model $\mathcal{H}_4$ and the reference Hamiltonian $\mathcal{H}^{\text{ref}}_4$. (b) The BdG spectrum $E$, where the red points correspond to the zero-energy eigenstates. The inset shows the spatial profile of one of the zero-energy eigenstates. The parity data of $\mathcal{H}_4$ and the figure in (b) are adapted from Ref. \onlinecite{ModelRXZ} with permission.  
} 
\label{fig:egH4}
\end{figure}
Finally, we compute the indicators for another BdG tight-binding model with momentum-dependent inversion operator proposed in Ref. \onlinecite{ModelRXZ}. The model $\mathcal{H}_4=h_{4,a}+h_{4,b}+h_{4,c}$ has the form 
\begin{align}
h_{4,a}=it&\sum_{{\bf{R}}}\beta_{{\bf{R}},AA}\alpha_{{\bf{R}},BB}+\beta_{{\bf{R}},AB}\alpha_{{\bf{R}}+{\bf{a}}_y,BA}\nonumber\\
&+\beta_{{\bf{R}},BA}\alpha_{{\bf{R}}+{\bf{a}}_x,AB}+\beta_{{\bf{R}},BB}\alpha_{{\bf{R}}+{\bf{a}}_x+{\bf{a}}_y,AA}\nonumber\\ 
\label{eq:h4a}
\end{align}
\begin{align}
h_{4,b}=im_x&\sum_{{\bf{R}}}\beta_{{\bf{R}},AA}\beta_{{\bf{R}},BA}+\alpha_{{\bf{R}},AA}\alpha_{{\bf{R}}-{\bf{a}}_x,BA}\nonumber\\
&+\beta_{{\bf{R}},AB}\beta_{{\bf{R}},BB}+\alpha_{{\bf{R}},AB}\alpha_{{\bf{R}}-{\bf{a}}_x,BB} 
\label{eq:h4b}
\end{align}
\begin{align}
h_{4,c}=im_y&\sum_{{\bf{R}}}\beta_{{\bf{R}},AA}\beta_{{\bf{R}},AB}+\alpha_{{\bf{R}},AA}\alpha_{{\bf{R}}-{\bf{a}}_y,AB}\nonumber\\
&+\beta_{{\bf{R}},BA}\beta_{{\bf{R}},BB}+\alpha_{{\bf{R}},BA}\alpha_{{\bf{R}}-{\bf{a}}_y,BB},  
\label{eq:h4c}
\end{align}
where the system has four sublattices $l=AA,BB,AB,BA$ per unit cell. Here, $\alpha_{{\bf{R}},l}$ and $\beta_{{\bf{R}},l}$ are again operators for the two flavors of Majorana fermions on the same sublattice $l$ in the same unit cell centered at ${\bf{R}}$, and ${\bf{a}}_x$, ${\bf{a}}_y$ are the lattice vectors. 
Similar to the previous model $\mathcal{H}_3$, here $\mathcal{H}_4$ also has a momentum-dependent inversion operator due to the sublattice structure of the model. The inversion operator has the form  
\begin{align}
\mathcal{I}_4({\bf{k}})={\text{diag}}&[1,e^{ik_x},e^{ik_y},e^{i(k_x+k_y)},\nonumber\\
&-1,-e^{ik_x},-e^{ik_y},-e^{i(k_x+k_y)}]      
\label{H1indicators}
\end{align}
in the basis $[c_{{\bf{k}},AA}$, $c_{{\bf{k}},BA}$, $c_{{\bf{k}},AB}$, $c_{{\bf{k}},BB}$, $c^{\dagger}_{-{\bf{k}},AA}$, $c^{\dagger}_{-{\bf{k}},BA}$, $c^{\dagger}_{-{\bf{k}},AB}$, $c^{\dagger}_{-{\bf{k}},BB}]$, where $c_{{\bf{k}},l}$ and $c^{\dagger}_{{\bf{k}},l}$ are the Fourier transforms of the complex fermion operators $c_{{\bf{R}},l}=(\alpha_{{\bf{R}},l}+i\beta_{{\bf{R}},l})/\sqrt{2}$ and $c^{\dagger}_{{\bf{R}},l}=(\alpha_{{\bf{R}},l}-i\beta_{{\bf{R}},l})/\sqrt{2}$. 
Importantly, it was numerically found in Ref. \onlinecite{ModelRXZ} that $\mathcal{H}_4$ hosts a higher-order phase with two Majoranas located at opposite corners [see Fig. \ref{fig:egH4}(b)].  

We now compute our indicators for $\mathcal{H}_4$. Given that $\mathcal{H}_4$ is an eight-band model, we choose the reference Hamiltonian to be 
\begin{align}
\mathcal{H}^{\text{ref}}_4({\bf{k}}) &= {\text{diag}}[\mathds{1}_4,-\mathds{1}_4].       
\label{H04}
\end{align}
In Fig. \ref{fig:egH4}, we display the parity data of the Fourier-transformed $\mathcal{H}_4$ computed in Ref. \onlinecite{ModelRXZ} and the parity data of $\mathcal{H}_4^{\text{ref}}$ we calculate using the momentum-dependent inversion operator $\mathcal{I}_4({\bf{k}})$. 
The resulting indicators we find are 
\begin{align}
&(\tilde{\kappa},\tilde{\nu}_x,\tilde{\nu}_y)=(2,0,0).  
\label{eq:indicatorsH4}
\end{align}
This indicates a strong higher-order phase with inversion-protected Majorana corner modes, which is consistent with the boundary modes numerically found in Ref. \onlinecite{ModelRXZ} [see Fig. \ref{fig:egH4}(b)].\\ 

\section{Summary and discussion}
In this work, we provide a faithful derivation for the symmetry indicators with a focus on the time-reversal invariant topological superconductors with wallpaper group $\bf{p_2}$ and an odd-parity superconducting order parameter. Specifically, our goal is to obtain indicators that can diagnose boundary features in the real space, and depend only on minimal set of essential symmetry data in the momentum space. We therefore take a double-pronged strategy that combines a real-space and a momentum-space classification schemes, namely the topological crystal approach where the boundary features become self-evident, and a twisted equivariant K group analysis where we find the informationally complete symmetry data of BdG bands for the classification purpose.

We begin with the topological crystal approach for the topological superconductors in this symmetry class. Importantly, besides the real-space classification and the corresponding boundary modes for each phase, we define a set of real-space invariants that characterizes the strong and weak phases. 

We then move to the momentum space and provide a detailed calculation of the K-theory classification by using the Atiyah-Hirzebruch spectral sequence. In particular, we show that the 0D topological invariants defined at the high-symmetry points alone serve as sufficient inputs for a complete set of topological invariants exhausting all phases classified by the K-theory. We also point out the importance of the reference Hamiltonian when defining the 0D topological invariants. Not including the parity data of the reference Hamiltonian (but only those of the BdG Hamiltonian of interest) could cause false negative and false positive results when diagnosing boundary Majoranas, especially for systems with  momentum-dependent inversion operators. 

The central result of our work is to perform a basis matching between our real-space topological invariants from the topological crystal approach and the symmetry indicators from our K-theory analysis. 
We emphasize that without this procedure, the physical meanings of the symmetry indicators are in fact ambiguous. This is because with the momentum-space approach alone, each symmetry indicator can correspond to an unknown mixture of strong and weak phases due to the freedom in choosing the basis for the homomorphism Eq.~\ref{eqn:f0}. 
Such an ambiguity could cause great confusions when one tries to apply the symmetry indicators to study realistic models or real materials. 
Our basis matching procedure disentangles the strong and weak phases in the indicator space so that the resulting indicators can provide a clear diagnosis for the boundary type, as we have demonstrated in several lattice models. 




Although there are many existing works on symmetry indicators for topological insulators and superconductors, this is the first time an indicator derivation is presented with a basis-matching procedure that guarantees the decomposition between strong and weak indicators. 
Without this basis-matching procedure we propose, the symmetry indicators cannot serve as a faithful diagnostic for boundary features. 
The principle of the derivation laid out in this work can be generalized to other symmetry classes. In particular, deriving indicators in the presence of other crystalline symmetries is an important future application of our formalism.  

\textit{Acknowledgement}-- We are grateful to Rui-Xing Zhang for helpful discussions, and to Mark Fischer, Hoi Chun Po, and Zhida Song for useful correspondence. This work is supported by the Laboratory for Physical Sciences. S.-J.H. acknowledges support from a JQI postdoctoral fellowship.

\bibliography{AHSS.bib}
 
\appendix
\section{Higher-order TSC as a stacking of two 2dTSCs}
\label{app:doubleTSC}
In this appendix, we demonstrate explicitly with a lattice model how stacking two copies of 2D class-DIII inversion-symmetric topological superconductors with counter-propagating Majorana edge modes (i.e. 2dTSC) gives rise to a 2D higher-order class-DIII superconductor with two Majorana Kramers pairs related by inversion. 

For each copy of the superconducting phases with edge modes, we choose to consider the lattice model $\mathcal{H}_1$ in Eq. \ref{eq:H1} (adapted from Ref. \onlinecite{WTe2HOTsc}) in the parameter regime where our indicators $\boldsymbol{\nu}=(\kappa,\nu_x,\nu_y)=(3,0,0)$ [see Eq. \ref{eq:indicatorsH1}]. 
As we described in Section VI.A, $\mathcal{H}_1$ 
obeys the time-reversal symmetry $\Theta_1=is_y\mathcal{K}$, $\textbf{k}\rightarrow-\textbf{k}$, the particle-hole symmetry $\Xi_1=\hat{\tau}_x\mathcal{K}$, $\textbf{k}\rightarrow-\textbf{k}$, and the inversion symmetry $\mathcal{I}_1=\hat{\tau}_z\otimes\hat{\rho}_z$,  $\textbf{k}\rightarrow-\textbf{k}$, where the odd-parity pairing demands $\{\mathcal{I}_1,\Xi_1\}=0$, and $\hat{\tau}$, $\hat{s}$, and $\hat{\rho}$ are Pauli matrices for particle and hole, spin $s=\uparrow,\downarrow$, and orbital $\rho=s,p_-$. 
Importantly, Ref. \onlinecite{WTe2HOTsc} has shown that by putting $\mathcal{H}_1$ on a rotational symmetric geometry and write the Hamiltonian in the polar coordinate $(r,\theta)$, one can analytically obtain the spin-up and spin-down Majorana edge modes 
\begin{align}
&\psi^{\uparrow/\downarrow}(r,\theta)=e^{-\frac{1}{\Delta}|r-R|}e^{il\theta}\left(\begin{array}{c}
0 \\e^{\pm i\frac{\theta}{2}} \\0 \\\mp ie^{\mp i\frac{\theta}{2}} \end{array}\right),  
\label{psi34up}
\end{align}
in the basis of $\hat{\tau}\otimes\hat{s}\otimes\hat{\rho}$. Here, $\sigma=1,2$ labels each of the two copies, $R$ is the radius of the geometry, $\Delta$ is superconducting gap magnitude [see Eq. \ref{eq:H1}], and $l$ is the orbital angular momentum taking half integers. 
The spin-up and spin-down edge modes are known to propagate in opposite directions with opposite angular momenta $2l$ and $-2l$, respectively\cite{WTe2HOTsc}.

We now stack two copies of 2dTSC by considering the block-diagonal model $\mathcal{H}=\hat{\sigma}_0\otimes\mathcal{H}_1$, where $\hat{\sigma}_i$ denotes the Pauli matrix in the copy basis $\sigma=1,2$. This \textit{double-TSC} model $\mathcal{H}$ clearly obeys the same symmetries as $\mathcal{H}_1$ since all the symmetries act on the $\hat{\sigma}$ space trivially. Moreover, in the absence of couplings between the two copies, we have two identicle sets of counter-propagating Majorana edge modes $\Psi^{\uparrow/\downarrow}_1 = ( (\psi^{\uparrow/\downarrow})^{T}, \boldsymbol{0}_{1 \times 4})^{T}$ and $\Psi^{\uparrow/\downarrow}_2 = ( \boldsymbol{0}_{1 \times 4}, (\psi^{\uparrow/\downarrow})^{T})^{T}$.

Next, we write down the lowest-order perturbations allowed by all the symmetries for the double-TSC model $\mathcal{H}$. Then by projecting these perturbations onto the two sets of counter-propagating edge modes, we will be able to examine if the perturbations can fully gap out the edge modes. 
Following what was done in Ref. \onlinecite{WTe2HOTsc} for a single copy of $\mathcal{H}_1$, we first write down the rotationally invariant perturbations $\mathcal{H}'_{\text{rot}}$ that remains invariant under the time-reversal symmetry $\Theta_1$, the particle-hole symmetry $\Xi_1$, the inversion symmetry $\mathcal{I}_1$, as well as the rotational symmetry $C_{\theta,1}$. Here, the rotational operation is given by $C_{\theta,1}=e^{-iJ_z\theta}$, $k_{\pm}\rightarrow e^{\mp i\theta}k_{\pm}$, where the angular momentum $J_z=\hat{\tau}_z\otimes \hat{s_z}\otimes\hat{\rho_z}/2$, and $k_{\pm}=ke^{\pm i\theta}$. $C_{\theta,1}$ also acts trivially on the copy space $\hat{\sigma}$. We find that the perturbations that obey $C_{\theta,1}\mathcal{H}'_{\text{rot}}C_{\theta,1}^{-1}=\mathcal{H}'_{\text{rot}}$ have the general form 
\begin{widetext}
\begin{align}
&\mathcal{H}'_{\text{rot}}(\textbf{k})=\nonumber\\
&\left(\begin{array}{cccccccccccccccc}
A_1 & A_2k_- & 0 & 0 & B_1k_- & B_2 & 0 & 0 & \alpha_1 & \alpha_2k_- & 0 & 0 & \beta_1k_- & \beta_2 & 0 & 0 \\
A_2^*k_+ & A_3 & 0 & 0 & -B_2 & B_3k_+ & 0 & 0 & \alpha_5k_+ & \alpha_3 & 0 & 0 & \beta_5 & \beta_3k_+ & 0 & 0 \\
0 & 0 & A_1 & -A_2^*k_+ & 0 & 0 & -B_1^*k_+ & B_2^* & 0 & 0 & \alpha_1^* & -\alpha_2^*k_+ & 0 & 0 & -\beta_1^*k_+ & \beta_2^* \\
0 & 0 & -A_2k_- & A_3 & 0 & 0 & -B_2^* & -B_3^*k_- & 0 & 0 & -\alpha_5^*k_- & \alpha_3^* & 0 & 0 & \beta_5^* & -\beta_3^*k_- \\
B_1^*k_+ & -B_2^* & 0 & 0 & -A_1 & A_2^*k_+ & 0 & 0 & \beta_1^*k_+ & -\beta_2^* & 0 & 0 & -\alpha_1^* & \alpha_2^*k_+ & 0 & 0 \\
B_2^* & B_3^*k_- & 0 & 0 & A_2k_- & -A_3 & 0 & 0 & -\beta_5^* & \beta_3^*k_- & 0 & 0 & \alpha_5^*k_- & -\alpha_3^* & 0 & 0 \\
0 & 0 & -B_1k_- & -B_2 & 0 & 0 & -A_1 & -A_2k_- & 0 & 0 & -\beta_1k_- & -\beta_2 & 0 & 0 & -\alpha_1 & -\alpha_2k_- \\
0 & 0 & B_2 & -B_3k_+ & 0 & 0 & -A_2^*k_+ & -A_3 & 0 & 0 & -\beta_5 & -\beta_3k_+ & 0 & 0 & -\alpha_5k_+ & -\alpha_3 \\
\alpha_1^* & \alpha_5^*k_- & 0 & 0 & \beta_1k_- & -\beta_5 & 0 & 0 & a_1 & a_2k_- & 0 & 0 & b_1k_- & b_2 & 0 & 0 \\
\alpha_2^*k_+ & \alpha_3^* & 0 & 0 & -\beta_2 & \beta_3k_+ & 0 & 0 & a_2^*k_+ & a_3 & 0 & 0 & -b_2 & b_3k_+ & 0 & 0\\
0 & 0 & \alpha_1 & -\alpha_5k_+ & 0 & 0 & -\beta_1^*k_+ & -\beta_5^* & 0 & 0 & a_1 & -a_2^*k_+ & 0 & 0 & -b_1^*k_+ & b_2^* \\
0 & 0 & -\alpha_2k_- & \alpha_3 & 0 & 0 & -\beta_2^* & -\beta_3^*k_- & 0 & 0 & -a_2k_- & a_3 & 0 & 0 & -b_2^* & -b_3^*k_- \\
\beta_1^*k_+ & \beta_5^* & 0 & 0 & -\alpha_1 & \alpha_5k_+ & 0 & 0 & b_1^*k_+ & -b_2^* & 0 & 0 & -a_1 & a_2^*k_+ & 0 & 0 \\
\beta_2^* & \beta_3^*k_- & 0 & 0 & \alpha_2k_- & -\alpha_3 & 0 & 0 & b_2^* & b_3^*k_- & 0 & 0 & a_2k_- & -a_3 & 0 & 0 \\
0 & 0 & -\beta_1k_- & \beta_5 & 0 & 0 & -\alpha_1^* & -\alpha_5^*k_- & 0 & 0 & -b_1k_- & -b_2 & 0 & 0 & -a_1 & -a_2k_- \\
0 & 0 & \beta_2 & -\beta_3k_+ & 0 & 0 & -\alpha_2^*k_+ & -\alpha_3^* & 0 & 0 & b_2 & -b_3k_+ & 0 & 0 & -a_2^*k_+ & -a_3 \\
\end{array}\right).  
\label{eq:Hrot}
\end{align}
\end{widetext}
Here, $\mathcal{H}'_{\text{rot}}(\textbf{k})$ is written in the basis of $\hat{\sigma}\otimes\hat{\tau}\otimes \hat{s}\otimes\hat{\rho}$, and $A_i$, $B_i$, $a_i$, $b_i$, $\alpha_i$, and $\beta_i$ for $i=1,2,3$ are free parameters.  
Since it is clear from Eq. \ref{eq:Hrot} that the rotational invariant perturbations in $\mathcal{H}'_{\text{rot}}(\textbf{k})$ do not couple spin-up and spin-down states, the two sets of helical edge modes from the two copies remain gapless under $\mathcal{H}'_{\text{rot}}(\textbf{k})$. 

We then examine the rotational-breaking perturbations. Here we consider only the lowest-order terms, which have no spatial dependence. After performing a similar procedure without the rotational symmetry, we find that there are only two non-vanishing symmetry-allowed terms that couple counter-propagating edge modes from different copies 
\begin{align}
&\mathcal{H}'_a=-\lambda_a\hat{\sigma}_y\otimes\hat{\tau}_z\otimes\hat{s_y}\otimes(\hat{\rho_0}-\hat{\rho_z})/2\nonumber\\
&\mathcal{H}'_b=-\lambda_b\hat{\sigma}_y\otimes\hat{\tau}_0\otimes\hat{s_x}\otimes(\hat{\rho_0}-\hat{\rho_z})/2. 
\label{eq:Hpertf}
\end{align}
By projecting these perturbations onto the edge modes $\Psi_{1}^{s}(r,\theta)$ and $\Psi_{2}^{s'}(r,\theta)$, we find 
their corresponding amplitudes after projection to be 
\begin{align}
&\int dr \Psi_{1}^{s\dagger}(r,\theta)(\mathcal{H}'_{a}+\mathcal{H}'_{b})\Psi^{\bar{s}}_{2}(r,\theta)
\propto\lambda_a\cos\theta+\lambda_b\sin\theta. 
\label{eq:Hpertf}
\end{align}
Since this back-scattering term $\lambda_a\cos\theta+\lambda_b\sin\theta$ between the counter-propagating edge modes has opposite signs at any $\theta$ and $\theta+\pi$, it has to vanish at some angle $\theta_0=\tan^{-1}(-\frac{\lambda_a}{\lambda_b})$ and $\theta_0+\pi$.
In other words, even at the lowest order, the rotational breaking perturbations fail to fully gap out the helical Majorana edge modes from the two copies. There will always be at least two `leftover' zero-dimensional zero-energy Kramer's pairs located at $\theta_0$ and $\theta_0+\pi$ when the double-TSC model $\mathcal{H}$ is placed on an open geometry. 
Importantly, the two Majorana pairs are related by inversion and can annihilate each other only when inversion symmetry is broken.
The specific value of $\theta_0$ is given by the microscopics, and we expect these two Majorana Kramers pairs to be trapped at the opposite corners of the considered geometry. 
This resulting higher-order state from our double-TSC model is therefore equivalent to a single $d_b=1$ topological crystal with $\boldsymbol{\nu}=(2,0,0)$, which has a block configuration that supports Majorana corner modes [see Fig. \ref{fig:model_HO}(a)].

\section{Topological phenomena interpretation of the first differentials}
\label{app:1std}

There are first differentials that only involve the diagonal and upper off-diagonal entries in the $E_1$-page table, and therefore have a clear physical meaning in the topological phenomena interpretation. In this appendix, we perform computation based on this interpretation for some entries in the $E_2$-page table. By comparing these results with our calculation using the representation interpretation in the main text, we find that the two interpretation lead to the same results as expected.  

Here we take $d_{1}^{0,-3}$ as an example. The mapping $d_{1}^{0,-3}$ describes a band inversion or chemical potential shift at a high symmetry point that creates two gapless points on adjacent high-symmetry lines. This process can be captured by the following minimal model:
\begin{equation}
H_{d_{1}^{0,-3}} = ( \boldsymbol{k}^{2} - \mu) \rho_{z} \otimes \sigma_{0} \otimes \tau_{z} + v k_{y} \rho_{y} \otimes \sigma_{0} \otimes \tau_{z}
\label{eqn:Hd1}
\end{equation}
Here $\tau$, $\sigma$, and $\rho$ are Pauli matrices for particle and hole, spin, and orbital degrees of freedom. The Hamiltonian Eq.~\ref{eqn:Hd1} obeys the time-reversal symmetry $\Theta=i \sigma_y\mathcal{K}$, $\textbf{k}\rightarrow-\textbf{k}$, the particle-hole symmetry $\Xi=\tau_x\mathcal{K}$, $\textbf{k}\rightarrow-\textbf{k}$, and the inversion symmetry $\mathcal{P}$ $=\tau_z\otimes\rho_z$,  $\textbf{k}\rightarrow-\textbf{k}$, where the odd-parity pairing demands $\{\mathcal{P},\Xi\}=0$. 
The band inversion can be realized by tuning the chemical potential $\mu$ from $\mu<0$ to $\mu>0$, and there will be two Dirac points on adjacent $1$-cells. Importantly, these Dirac points are not protected and can be gapped out by a symmetry-allowed $p$-wave pairing term $H_{\Delta_{x}} = \Delta k_{x}  \rho_{0} \otimes \sigma_{z} \otimes \tau_{x}$. 
This is consistent with $d_{1}^{0,-3}$ being a trivial map, which leads to $E_{2}^{0,-3}=\text{Ker} (d_{1}^{0,-3})=E_1^{0,-3}=\z\times\z\times\z\times\z$. 
This is consistent with the what we found in the main text [see Table \ref{table:E2}].

\end{document}